\documentclass[11pt]{article}
\pdfoutput=1
\usepackage[centertags]{amsmath}
\usepackage{latexsym}
\usepackage{mathtools}
\usepackage{amssymb}
\usepackage{graphicx}
\usepackage{epsfig}
\usepackage{ulem}
\usepackage[english]{babel}
\usepackage{array}
\usepackage{amsthm}
\usepackage{latexsym}
\usepackage[mathcal]{euscript}
\pdfoutput=1
\usepackage{epsfig}

 \usepackage{jheppub}
 \usepackage{hyperref}

\usepackage{subfig}
\usepackage{psfrag}

\usepackage[latin1]{inputenc}
\usepackage{float}
\usepackage{cancel}
\usepackage{mathrsfs}
\usepackage{amssymb}
\usepackage{amsfonts}
\usepackage{amsmath}
\usepackage{slashed}
\usepackage{bm}
\usepackage{color}
\usepackage{appendix}
\newcommand{\be}{\begin{equation}}
\newcommand{\ee}{\end{equation}}
\newcommand{\bea}{\begin{eqnarray}}
\newcommand{\eea}{\end{eqnarray}}
\newcommand{\bwt}{\begin{widetext}}
\newcommand{\ewt}{\end{widetext}}

\newcommand{\bi}{\begin{itemize}}
\newcommand{\ei}{\end{itemize}}

\usepackage{setspace}

\definecolor{dgreen}{rgb}{0.,0.6,0.}

\begin{document}
%\doublespacing

\title{Higher-dimensional chaotic features and random matrix signatures following a local quench}

\author[a,b]{Dmitry S. Ageev}
\author[c]{and Jacob Sonnenschein}
\affiliation[a]{
  Steklov
Mathematical Institute, Russian Academy of Sciences, Gubkin str. 8, 119991 Moscow, Russia }
\affiliation[b]{
 Institute for Theoretical and Mathematical Physics, Lomonosov Moscow State University, 119991 
Moscow, Russia}

\affiliation[c]{The Raymond and Beverly Sackler School of Physics and Astronomy, \\
Tel Aviv University, Ramat Aviv 69978, Tel Aviv, Israel}
\emailAdd{ageev@mi-ras.ru}
\emailAdd{cobi@tauex.tau.ac.il}

\abstract
{
We study the multidimensional erratic structure of correlation functions produced by local operator quenches in finite-volume free massive scalar field theory in dimensions 2 and 3.  The basic observable is the subtracted equal-time two-point function in the locally excited state and its spatiotemporal patterns of extrema.  We analyze these extrema by the multidimensional diagnostics recently introduced for chaotic scattering amplitudes and related problems: all-pair distance distributions, nearest-neighbor spacings, greedy-path spacing ratios, and the extrema form factor.  For the $1+1$-dimensional local quench we find that, in the regime of small Euclidean smearing, the fitted extremum statistics move close to the $\beta=1$ random-matrix benchmark, while increasing the smearing scale softens the effective repulsion and moves the distributions away from the GOE-like value.  For the $2+1$-dimensional local quench we find that the nearest-neighbor statistics of the refined extrema are close to, or above, the $\beta=1$ benchmark, and the greedy-path ratio statistics are described by even larger effective $\beta$ values. Finally we studied the all-pair extrema spatial form factor and found that, in the one-, two-, and three-dimensional cases, its main structure is controlled by the corresponding uniform interval, rectangle, or cuboid geometry of the extrema cloud and found the dip-ram-plateau structure in the last two cases. Thus the form factor provides a complementary global diagnostic of how the extrema fill their effective metric support, while the genuinely nontrivial local and mesoscopic organization is carried by the nearest-neighbor and greedy-path statistics.
} 

\maketitle

\newpage

\section {Introduction}

Chaotic quantum systems are often recognized not through a single large number or a single sharp observable, but through the way in which otherwise ordinary functions acquire an erratic dependence on their natural variables.  In spectral problems this idea is made precise by the statistics of energy levels \cite{Berry:1977wk,Bohigas:1983er,Mehta:book,Haake:2010fgh}, while in scattering problems it can be expressed through the oscillatory dependence of amplitudes on kinematical invariants, and in time-dependent field-theoretic problems it may instead appear in the intricate spacetime structure of correlation functions following a perturbation (see  \cite{Calabrese:2007mt,Calabrese:2007rg,Nozaki:2013wia,Nozaki:2014hna} for non-chaotic behaviour in conformal field theory).  A useful diagnostic of such behavior should therefore not be tied too rigidly to one preferred observable, especially in situations where the usual spectral or correlation-function probes either do not exist in a convenient form or do not isolate the most irregular part of the dynamics.

A recent proposal in this direction was made in \cite{Bianchi:2022mhs,Bianchi:2023uby}, where the erratic behavior of string scattering amplitudes was quantified by studying the extrema of the relevant functions rather than the functions themselves (for the studies of amplitudes of this kind also see   \cite{Rosenhaus:2020tmv,Gross:2021gsj,Rosenhaus:2021xhm,Hashimoto:2022bll,Savic:2024ock}).  The virtue of this approach is that the extrema form a geometric and statistical object which can be compared to random-matrix expectations without requiring the original problem to be a conventional one-dimensional spectral problem.  In particular, the distributions of spacings between neighboring extrema, the ratios of such spacings, and the associated ``scattering form factor'' \cite{Bianchi:2024fsi} provide a map between the irregular geometry of the amplitude and the structures familiar from random matrix theory \cite{Mehta:book,Forrester:book}.  The map is not a literal identification of the amplitude with a random matrix spectrum; rather, it is a way of translating the visible irregularity of a function into statistical data that can be compared across different physical systems.

This point of view has already proved useful outside its original scattering-amplitude context and generalized on higher-dimensions.  
In 
\cite{Bianchi:2025kna}  the one dimensional chaotic behavior  was generalized to higher dimensional models, namely, those that depend on several kinematical variables.  It was  examined  in the context of the pinball scattering  as well as a toy model of the electric potential produced by a set of charges located randomly.

  In \cite{Ageev:2025yiq} it was applied to local and global quenches  of a free massive scalar field theory in two dimensions \cite{Ageev:2022kpm}, where the one-dimensional dependence of certain observables on time, space, or Euclidean separation gives rise to extrema whose spacings can be analyzed in close analogy with the spacing statistics of erratic amplitudes.  A similar structure was found for two-point correlators of these theories when they are regarded as functions of a single variable.  These results suggest that the extrema of quenched correlation functions carry information about the organization of the post-quench interference pattern, even in theories which are free and therefore not chaotic in the usual many-body sense. In \cite{Ageev:2025iiy} this setup has been generalized to the case of AdS
 and the chaotic behavoiur has been revealed in the behaviour of the boundary observables.

There is also a closely related, but conceptually distinct, line of work in which the quench itself is used as a probe of chaotic dynamics.  In spin chains and conformal field theories, out-of-time-order correlators after a quench have been used to diagnose information scrambling, to distinguish chaotic from integrable time evolution, and in some cases to identify equilibrium or dynamical phase transitions \cite{Heyl:2018otc,Alba:2019scrambling,Das:2021otoc}.  In holographic systems, sudden energy injection gives a particularly sharp realization of this idea, since the corresponding Vaidya geometries allow one to follow the time-dependent Lyapunov growth and butterfly cones of post-quench OTOCs \cite{Balasubramanian:2019quench}.  More recently, local quenches in two-dimensional conformal field theories have also been reformulated in Krylov space, where spread complexity and Krylov entropy provide another language for the growth of a localized perturbation \cite{Caputa:2025krylov}.  The present paper is in the same broad spirit, but uses a different diagnostic: instead of extracting chaos from an OTOC, a Loschmidt echo, or Krylov complexity, we study the multidimensional geometry of extrema of ordinary quenched two-point functions.

The next natural step is to abandon the restriction to one-dimensional scans.  A local quench on a finite interval does not produce a function of time alone \cite{Calabrese:2007rg,Nozaki:2014hna,Ageev:2025yiq}.  It produces a spacetime pattern, because the signal emitted by the local insertion propagates through the finite volume, reflects from the boundaries, and interferes with itself after repeated returns.  The resulting subtracted two-point function depends simultaneously on the observation point and on time, and its extrema therefore form a cloud in the two-dimensional $(x,t)$ plane rather than an ordered sequence on a line.  In a higher-dimensional box the same construction leads to an extrema cloud in $(x,y,t)$.  

The appropriate multidimensional framework was introduced in \cite{Bianchi:2025kna}, where the one-dimensional notion of erratic behavior was generalized to functions of several kinematical variables.  There the method was tested in the context of pinball scattering and also in a toy model given by the electric potential produced by charges placed randomly on a plane.  Three complementary diagnostics were introduced for the resulting cloud of extrema: the all-pair distance distribution, which probes the global support of the cloud; the nearest-neighbor spacing distribution, which probes the local repulsion or clustering of extrema; and the greedy-path spacing and ratio statistics, which impose a concrete ordering on the cloud and thereby probe a mesoscopic organization which is invisible in purely local data.  Somewhat unexpectedly, the best phenomenological descriptions of the latter statistics were not always the Gaussian-$\beta$ ensembles \cite{Dumitriu:2002,Forrester:book} or the ABGVV formula \cite{ABGVV,Atas:2013dis}, but rather a logistic distribution for the nearest-neighbor spacings and an ordinary Beta distribution for the folded spacing ratios.

The purpose of the present paper is to apply this multidimensional extremum-statistics program to quenched correlation functions.  We focus first on a local operator quench of a free massive scalar field on a finite interval, where the observable of interest is the vacuum-subtracted equal-time two-point function with one endpoint pinned at the quench location.  This observable is exactly computable by the interval mode expansion, but its dependence on $(x,t)$ is highly oscillatory because of the interplay between the Euclidean smearing of the local insertion, the finite-volume spectrum, and the repeated boundary reflections of the wave packet.  Instead of attempting to characterize this complicated function directly, we extract its maxima and minima in a prescribed spacetime window and treat the resulting extrema as the fundamental statistical object.  We then repeat the analysis for the natural $2+1$-dimensional generalization, where the interval is replaced by a rectangular spatial box and the extrema form a three-dimensional cloud.

The analysis proceeds in several steps.  First we extract the refined extrema of the local-quench correlator and separate maxima, minima, and the combined cloud of all extrema.  We then compute all-pair distances, nearest-neighbor spacings, greedy-path spacings, and greedy-path spacing ratios for each cloud.  The corresponding probability distributions are compared to the Gaussian-$\beta$ spacing family and to the ABGVV-type ratio family, but also to the logistic and ordinary Beta distributions which were found to be effective in the multidimensional examples of \cite{Bianchi:2025kna}.  Finally, we define and compute an all-pair extremum spatial form factor, denoted by $E_xFF$, which is the Fourier transform of the empirical all-pair distance distribution and therefore serves as the point-cloud analogue of a spectral or scattering form factor \cite{Cotler:2016fpe,Liu:2018hlr,Bianchi:2024fsi,Massaro:2024xfd}.  A central question throughout the paper is whether the extrema of quenched correlators reproduce the usual one-dimensional random-matrix laws, or whether they instead realize the broader multidimensional pattern in which random-matrix-inspired fits coexist with logistic and Beta laws controlled by the geometry of extrema clouds.

$\,$

The paper is organized as follows.  In section 2 we review the local operator quench of a free massive scalar field on an interval, beginning with the action, Hamiltonian, exact open boundary conditions, and interval mode expansion, and then deriving the subtracted two-point observable whose extrema will be studied.  In section \ref{ExpO} we introduce the extremum-point observables, including all-pair correlations, nearest-neighbor spacings, greedy-path spacing ratios, and the unfolding issues which are special to multidimensional point clouds.  In section \ref{NuRe} we present the numerical results for the interval local quench, with separate discussions of the global all-pair geometry, the nearest-neighbor repulsion, the greedy-path ratio statistics, and the dependence of the fitted effective parameters on the Euclidean smearing scale.  In section \ref{3d} we extend the construction to the $2+1$-dimensional local quench in a rectangular box and analyze the resulting three-dimensional extrema cloud. Finally we introduce the all-pair extremum spatial form factor and compare the one-, two-, and three-dimensional cases.  We conclude with a summary of the main lessons and with several open questions, while Appendix A records the numerical implementation, the refinement of extrema, the metric conventions, and the convergence issues relevant for the computations.

\section{Local operator quench of a free scalar on an interval}

\subsection{Action, Hamiltonian, and  open boundary conditions}

We consider a real scalar field $\phi(t,x)$ of mass $m$ on the strip $\mathbb{R}_t \times [0,L]_x$, with action
\begin{equation}
S = \frac{1}{2}\int dt \int_0^L dx \, \left[ (\partial_t \phi)^2 - (\partial_x \phi)^2 - m^2 \phi^2 \right].
\end{equation}
Throughout the present work  we consider finite spatial regions (namely spatial interval here) and the boundary conditions are imposed  at the two endpoints. At each endpoint we choose either Dirichlet or Neumann boundary conditions (with N or D standing for each notation of each choice in what follows). Thus, at $x=0$ and $x=L$ we impose
\begin{gather}
\phi(t,0) = 0 \qquad \text{or} \qquad \partial_x \phi(t,0)=0, \\
\phi(t,L) = 0 \qquad \text{or} \qquad \partial_x \phi(t,L)=0.
\end{gather}
 Denote $u_a(x)$  an orthonormal basis of eigenfunctions of the spatial Laplacian, normalized on $[0,L]$, i.e.
\begin{equation}
-\partial_x^2 u_a(x) = k_a^2 u_a(x), \qquad \int_0^L dx \, u_a(x)u_b(x) = \delta_{ab},
\end{equation}
with frequencies defined via
\begin{equation}
\omega_a = \sqrt{m^2 + k_a^2}.
\end{equation}
For the four exact interval boundary-condition sectors that are relevant for the present paper one has the  spectra
\begin{gather}
\text{DD:} \qquad u_n(x)=\sqrt{\frac{2}{L}}\sin\frac{n\pi x}{L}, \qquad k_n=\frac{n\pi}{L}, \qquad n=1,2,\dots, \\
\text{NN:} \qquad u_0(x)=\frac{1}{\sqrt{L}}, \qquad u_n(x)=\sqrt{\frac{2}{L}}\cos\frac{n\pi x}{L}, \qquad k_n=\frac{n\pi}{L}, \qquad n=1,2,\dots, \\
\text{DN:} \qquad u_n(x)=\sqrt{\frac{2}{L}}\sin\frac{(n+\frac{1}{2})\pi x}{L}, \qquad k_n=\frac{(n+\frac{1}{2})\pi}{L}, \qquad n=0,1,\dots, \\
\text{ND:} \qquad u_n(x)=\sqrt{\frac{2}{L}}\cos\frac{(n+\frac{1}{2})\pi x}{L}, \qquad k_n=\frac{(n+\frac{1}{2})\pi}{L}, \qquad n=0,1,\dots.
\end{gather}
The field operator is expanded in their terms as
\begin{equation}
\phi(t,x)=\sum_a \frac{1}{\sqrt{2\omega_a}}\left( a_a e^{-i\omega_a t}+a_a^\dagger e^{i\omega_a t} \right)u_a(x),
\end{equation}
with the standard canonical commutation relations. 

$\,$

The Euclidean two-point correlation function  on the interval  at inverse temperature $\beta_T$  is written in terms of $u_a(x)$ as 
\begin{equation}
G_{\beta_T}(\tau;x,y)=\frac{1}{A}\sum_a \frac{u_a(x)u_a(y)}{2\omega_a\left(1-e^{-\beta_T \omega_a}\right)}
\left( e^{-\omega_a \tau}+e^{-\omega_a(\beta_T-\tau)} \right),
\end{equation}
 where  we denote Euclidean time separation $\tau$,  spatial points $x,y \in [0,L]$ and
 $A$ is an overall normalization constant.  In the regime used in the numerical examples below we take $\beta_T$ to be very large, so that the calculation is effectively in the vacuum sector, but it is useful to retain the thermal notation because the Euclidean regulator of the local quench fits naturally into the same formalism.

\subsection{Locally excited state and the observable of interest}

The quench state is produced by the local insertion of the field operator at a point $x_q \in (0,L)$, dressed by the Euclidean smearing $e^{-\alpha H}$. At time $t=0$ we define
\begin{equation}
|\Psi_\alpha(0)\rangle = \frac{e^{-\alpha H}\phi(x_q)|0\rangle}{\bigl\langle 0 \bigl| \phi(x_q)e^{-2\alpha H}\phi(x_q)\bigr|0 \bigr\rangle^{1/2}},
\end{equation}
while at later times we evolve with the exact Hamiltonian,
\begin{equation}
|\Psi_\alpha(t)\rangle = e^{-iHt}|\Psi_\alpha(0)\rangle.
\end{equation}
In what follows we will focus on  the equal-time two-point  observable with one endpoint pinned at the quench location and with the subtraction of vacuum correlation part
\begin{equation}
\mathcal{G}_{\rm loc}(t,x)=
\langle \Psi_\alpha(t)|\phi(t,x)\phi(t,x_q)|\Psi_\alpha(t)\rangle
-
\langle 0|\phi(t,x)\phi(t,x_q)|0\rangle.
\end{equation}
Because the theory is Gaussian, Wick contraction reduces this quantity to a simple bilinear expression in the analytically continued two-point function. A convenient form for numerical calculations is
\begin{equation}\label{eq:loca}
\mathcal{G}_{\rm loc}(t,x)=
\frac{2\,\mathrm{Re}\!\left[
G_{\beta_T}(\alpha+i t;x,x_q)^\ast G_{\beta_T}(\alpha+i t;x_q,x_q)
\right]}
{G_{\beta_T}(2\alpha;x_q,x_q)}.
\end{equation}

\begin{figure}[t!]
    \centering
    \includegraphics[width=1.1\linewidth]{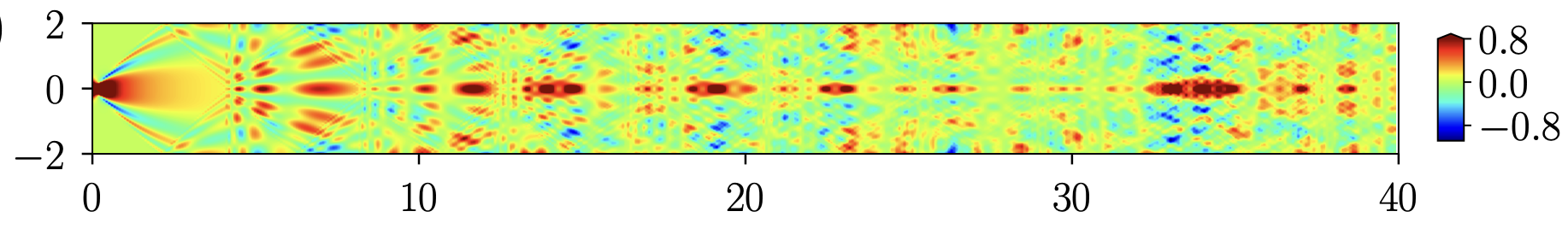}
    \caption{The two-point correlation function defined via \eqref{eq:loca} for the strip size $L=4$, mass $m=10$, $\alpha=0.05$.}
    \label{fig:figG}
\end{figure}

This function is oscillatory in time producing complicated spatiotemporal pattern reflecting the interplay between the localized insertion, the finite-volume mode structure, and the repeated propagation and reflection of the wave packet from the two boundaries. On the interval, the repeated return of signal from the boundaries produces an interference pattern in $(x,t)$ which is sufficiently irregular that its extremum set becomes a meaningful statistical object in its own right. That extremum set, rather than the full correlator, is the arena in which the rest of the paper is formulated. One can see the behaviour of typical subtracted correlated defined via \eqref{eq:loca}  in Fig. \ref{fig:figG} for a patricular set of parameters.

\section{Extremum-point observables, comparison curves, and unfolding}\label{ExpO}

\subsection{Extremum sets and the analysis window}

Let $\Omega \subset [0,L]\times \mathbb{R}_+$, i.e. the  rectangle in $(x,t)$ and  excise a small core around the insertion point in order to suppress the ultraviolet region in which the short-distance regularization dominates the geometry of the correlator. Therefore define
\begin{equation}
\Omega = \left\{ (x,t)\, \big|\, x_{\rm min}\le x \le x_{\rm max}, \quad
t_{\rm min}\le t \le t_{\rm max}, \quad
(x-x_q)^2+t^2 > r_{\rm core}^2 \right\}.
\end{equation}
Within this region we define the sets of maxima and minima,
\begin{gather}
\mathcal{E}_+ = \left\{ p_i^+ = (x_i^+,t_i^+) \in \Omega \, \big| \, p_i^+ \text{ is a local maximum of } \mathcal{G}_{\rm loc}\right\}, \\
\mathcal{E}_- = \left\{ p_i^- = (x_i^-,t_i^-) \in \Omega \, \big| \, p_i^- \text{ is a local minimum of } \mathcal{G}_{\rm loc}\right\},
\end{gather}
and  also consider the combined cloud of all extrema
\begin{equation}
\mathcal{E} = \mathcal{E}_+ \cup \mathcal{E}_-.
\end{equation}
Separating maxima from minima is not merely cosmetic. The two sets need not have the same density, and in a general oscillatory pattern they can probe slightly different parts of the local properties of the underlying field in the analysis that follows. At the same time, the combined cloud is often the object most naturally compared to a generic two-dimensional point process. For that reason the later numerical analysis is carried out on all of these sets.  Now let us subsequently introduce different measures \cite{Bianchi:2025kna} which we will apply to different sets of  $\mathcal{E}$ for correlation function $\mathcal{G}_{\rm loc}$ .

\subsection{All pair correlations}

The first measure is the simplest and also the most global. Given a point cloud $\mathcal{E}=\{p_i\}_{i=1}^N$ in two dimensions, we form all pairwise distances
\begin{equation}
d_{ij} = |p_i-p_j|, \qquad 1\le i<j\le N.
\end{equation}
This produces $N(N-1)/2$ distances averaging over a large amount of information and the principal virtue of this measure is that it does not require any ordering prescription. The principal drawback is that it mixes local geometry, global geometry, density variation, and boundary effects in a single distribution. For that reason it should not be thought of as an analogue of nearest-neighbor level-spacing statistics. Rather, it is a global diagnostic for whether the point cloud, once expressed in the chosen metric coordinates, resembles a nearly uniform distribution in a rectangle or instead retains strong remnants of inhomogeneity.

For a uniform point cloud in a rectangle of sides $a$ and $b$, the pair-distance distribution is known analytically. A useful representation is
\begin{equation}\label{eq: poisson-allpair}
p_{\rm rect}(r;a,b)=
\frac{4r}{a^2 b^2}
\int_{\theta_-(r)}^{\theta_+(r)}
(a-r\cos\theta)(b-r\sin\theta)\, d\theta,
\end{equation}
where the angular limits enforce the condition that the displacement vector of length $r$ remain inside the rectangle. In the numerical analysis we compare the histogram of normalized pair distances $r/\langle r\rangle$ to this exact reference curve. When the agreement is good, one learns that the cloud is already close to globally uniform in the chosen metric variables. When the agreement fails, one learns either that the density has not been properly flattened or that the point cloud possesses genuine correlations beyond those of a uniform random set.

\subsection{Nearest-neighbor spacings}

The second measure focuses on local repulsion. For each point $p_i$ in the cloud we define the nearest-neighbor spacing
\begin{equation}
\delta_i^{\rm NN} = \min_{j\ne i} |p_i-p_j|.
\end{equation}
After normalizing by the sample mean,
\begin{equation}
s_i = \frac{\delta_i^{\rm NN}}{\langle \delta^{\rm NN} \rangle},
\end{equation}
we obtain a distribution that may be compared to the one-parameter repulsive reference families. The first such family is the Gaussian $\beta$-ensemble spacing surmise
\begin{equation}
p_\beta(s)=C_\beta s^\beta e^{-\gamma_\beta s^2},
\end{equation}
with
\begin{gather}
C_\beta = 2 \frac{\Gamma\!\left(\frac{\beta+2}{2}\right)^{\beta+1}}
{\Gamma\!\left(\frac{\beta+1}{2}\right)^{\beta+2}}, \,\,\,\,\,\,\,\,
\gamma_\beta = 
\left(
\frac{\Gamma\!\left(\frac{\beta+2}{2}\right)}
{\Gamma\!\left(\frac{\beta+1}{2}\right)}
\right)^2.
\end{gather}
In one-dimensional spectral statistics the case $\beta=1$ is the familiar GOE surmise. In the present two-dimensional problem it should be interpreted more cautiously, because the same small-$s$ linear repulsion can arise for geometric reasons that have nothing to do with a one-dimensional random-matrix spectrum.

A second comparison curve, motivated by the multi-dimensional chaos literature\cite{Bianchi:2025kna}, is the logistic distribution
\begin{equation}
p_{\rm log}(s;\mu,\sigma)=
\frac{e^{-(s-\mu)/\sigma}}
{\sigma \left(1+e^{-(s-\mu)/\sigma}\right)^2}.
\end{equation}
In the numerical fits displayed below the normalization of the spacings sets $\mu=1$, so that only the width parameter $\sigma$ is fitted. The reason to include this curve is empirical rather than doctrinal: once the data arise from extrema of a genuinely two-dimensional function, a broad single-hump distribution centered near the mean spacing may describe the data more faithfully than a strict Wigner-like law.

\subsection{Greedy-path adjacent spacings and spacing ratios}

The third measure interpolates between the local and the ordered points of view. Since a generic two-dimensional point cloud comes with no canonical ordering, one needs an auxiliary rule if one wants to define successive spacings and their ratios. We follow the greedy-path method applied to the our extrema sets as introduced in \cite{Bianchi:2025kna}. Starting from a chosen initial point $p_{i_1}$, the next point $p_{i_2}$ is defined to be its nearest unvisited neighbor; then $p_{i_3}$ is defined to be the nearest unvisited neighbor of $p_{i_2}$; and the process continues until all points are exhausted. This produces a sequence of path spacings
\begin{equation}
\delta_n = |p_{i_{n+1}}-p_{i_n}|,
\end{equation}
and therefore a sequence of adjacent ratios
\begin{equation}
r_n = \frac{\delta_{n+1}}{\delta_n}.
\end{equation}
As in one-dimensional spacing-ratio studies, it is convenient to fold these ratios to the unit interval,
\begin{equation}
\widetilde r_n = \min(r_n,r_n^{-1}),
\end{equation}
so that $\widetilde r_n \in [0,1]$. The obvious advantage of this measure is that it uses ratios and therefore avoids a separate normalization of the spacings. The equally obvious drawback is that it depends on the ordering prescription and, in particular, on the chosen starting point. In practice, however, it probes a mesoscopic organization of the point cloud which is not visible in purely nearest-neighbor data and is therefore useful as an independent diagnostic.

Our first comparison curve for the folded ratios is the standard ABGVV-type $\beta$-family\cite{ABGVV},
\begin{equation}\label{Atas}
p_\beta(\widetilde r)=
2 \, \frac{1}{Z_\beta}
\frac{(\widetilde r + \widetilde r^2)^\beta}
{(1+\widetilde r+\widetilde r^2)^{1+\frac{3}{2}\beta}},
\end{equation}
where $Z_\beta$ is the normalization constant and the factor of $2$ folds the raw ratio distribution to the interval $[0,1]$. 

In \cite{Bianchi:2025kna}  it was found out that the pdf of the ratios of the spacings does not follow (\ref{Atas}) but rather a Beta distribution of the following form 

\begin{equation}
p_{\rm B}(\widetilde r;a,b)=
\frac{\Gamma(a+b)}{\Gamma(a)\Gamma(b)}
\, \widetilde r^{\,a-1}(1-\widetilde r)^{\,b-1}.
\end{equation}
When $a$ and $b$ are both slightly above unity, this family produces a broad plateau-like shape with a mild suppression near both endpoints, which is often a very accurate description of empirical path-ratio data in two-dimensional peak problems. Because of the  results of  \cite{Bianchi:2025kna}   we compare the histograms to the  ordinary Beta distribution.

\subsection{Why unfolding is indispensable and why it is subtle}

In one-dimensional spectral statistics the role of unfolding is to remove the smooth part of the density so that the mean local level spacing is constant. In two dimensions the same idea survives, but it immediately becomes more delicate because density flattening does not uniquely determine the metric geometry. If the extremum cloud has density $\rho(x,t)$, a two-dimensional unfolding seeks new coordinates $(u,v)$ such that the transformed density is constant. A natural construction is based on cumulative distributions,
\begin{gather}
u = \sqrt{N}\, F_x(x), \\
v = \sqrt{N}\, F_{t|x}(t|x),
\end{gather}
which is a Rosenblatt-type map. In the special case in which the two coordinates are independent, the conditional CDF factorizes and one recovers the one-dimensional unfoldings of $x$ and $t$. In general, however, different choices of ordering of the variables, different smoothing prescriptions, and different local-density estimators lead to different unfolded geometries. This is not a minor technical nuisance. A nonlinear unfolding can change the nearest-neighbor graph itself and can therefore change not only the numerical values of the spacings but also the path that enters the ratio statistic.

For that reason we distinguish sharply between two operations. The first is a purely global anisotropy correction, which we call axis rescaling,
\begin{equation}
x \mapsto \frac{x}{\ell_x}, \qquad t \mapsto \frac{t}{\ell_t},
\end{equation}
with $\ell_x$ and $\ell_t$ taken from the mean projected spacings along the two axes. This removes the trivial fact that the time window is much larger than the spatial one, but it does not flatten the density. The second is a genuine density unfolding, which attempts to make the point cloud approximately uniform in the transformed variables. The no-unfolding figures displayed in the present paper correspond to the first operation, namely axis rescaling without density flattening. This is already informative, but it does not by itself isolate universal local statistics.

\iffalse

There is a further caveat that is so important that it deserves to be stated explicitly in the body of the paper rather than hidden in a footnote. In two dimensions the nearest-neighbor spacing distribution of a homogeneous Poisson process is (\textcolor{red}{ HERE W})
\begin{equation}
P_{\rm Poi}(r)=2\pi \rho \, r\, e^{-\pi \rho r^2}.
\end{equation}
Since $\langle r\rangle = 1/(2\sqrt{\rho})$, the mean-normalized spacing $s=r/\langle r\rangle$ has density
\begin{equation}
p_{\rm Poi}(s)=\frac{\pi}{2}s\, e^{-\pi s^2/4}.
\end{equation}
This is exactly the Wigner surmise for the GOE spacing distribution. Therefore a good $\beta=1$ fit of the nearest-neighbor histogram in two dimensions is not, by itself, evidence for a random-matrix mechanism. It may simply express the geometry of a locally homogeneous point cloud. This observation does not invalidate the use of $\beta$-fits; rather, it forces one to interpret them as phenomenological descriptors unless they are supported by stability under unfolding and by comparison across several independent measures.
\fi
\section{Probability distributions of  spacings -Numerical results}\label{NuRe}

We now turn to the numerical data obtained from the interval local-quench correlator. The discussion in this section is intentionally detailed because the main point is not merely to record fitted parameters but to understand what each figure is actually telling us about the geometry of the extremum cloud. The datasets displayed below are generated from the exact interval-mode implementation with a local insertion at $x_q=L/2$, a large inverse temperature $\beta_T=10^4$ so that the calculation is effectively in the vacuum sector, mass $m=10$, Euclidean regulator $\alpha=0.05$, spatial interval $L=1$, cutoff $k_{\rm max}=5200$, spatial grid size $N_x=1000$, temporal grid size $N_t=30000$, time window $t\in[0.2,300]$, and an excluded core of radius $r_{\rm core}=0.08$ around the quench point. The figures shown here correspond to the axis-rescaled, not density-unfolded, analysis of the refined extremum sets. The raw and refined counts are $4545\to4015$ for maxima, $5370\to4787$ for minima, and $9915\to8802$ for the combined set. The labels in the figures indicate whether the cloud consists of maxima, minima, or all extrema. The discussion below is written so as to apply uniformly to any of the exact interval sectors once the figure is fixed.

\begin{figure}[t]
\centering
 \includegraphics[width=\textwidth]{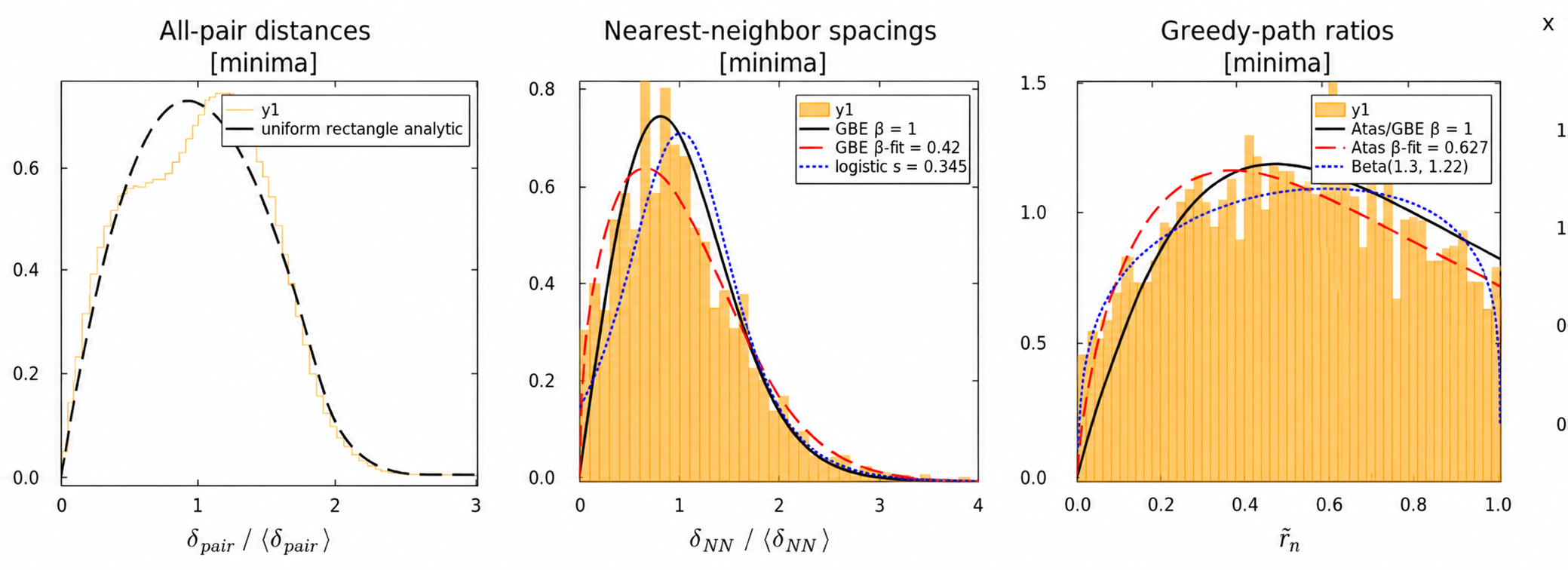}
\caption{Axis-rescaled, non-density-unfolded analysis of the refined minima cloud. The left panel shows the histogram of all pair distances together with the analytic uniform-rectangle reference curve \eqref{eq: poisson-allpair}. The middle panel shows the nearest-neighbor spacing distribution and its Gaussian-$\beta$ and logistic fits. The right panel shows the greedy-path ratio distribution together with the ABGR-type $\beta$ fit and the ordinary Beta fit.}
\label{fig:minima_axis}
\end{figure}

\begin{figure}[t]
\centering
 \includegraphics[width=\textwidth]{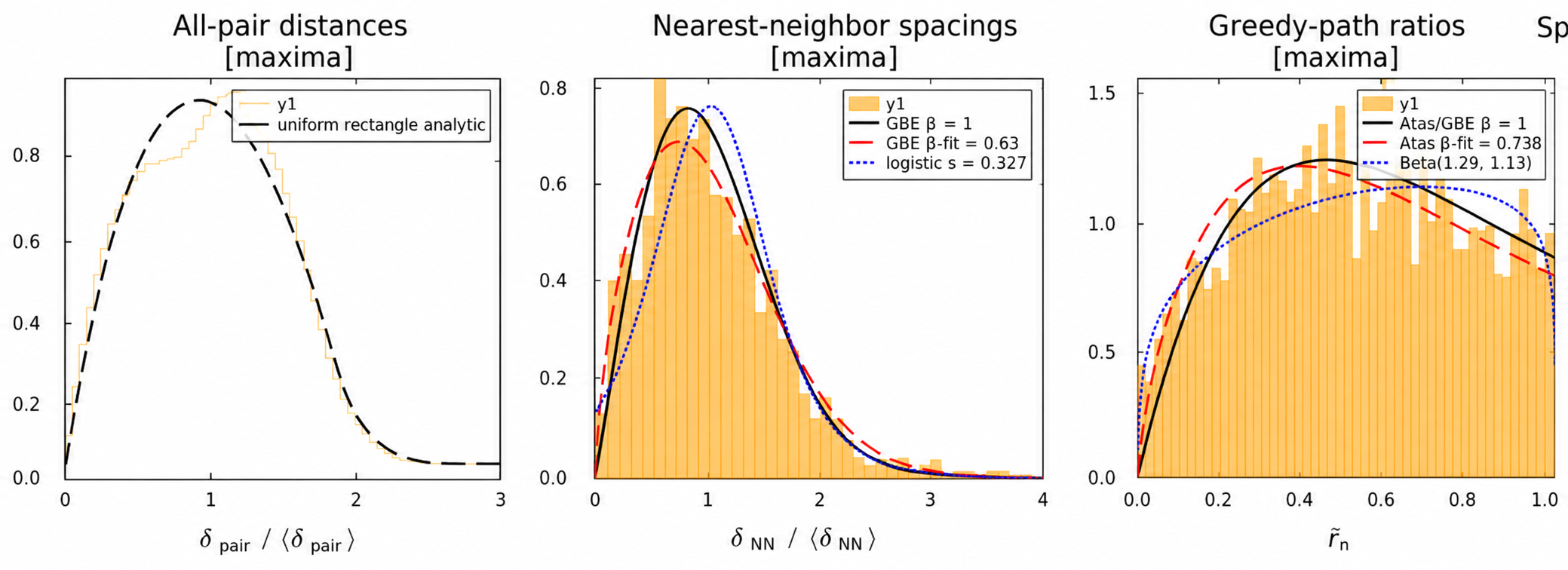}
\caption{Axis-rescaled, non-density-unfolded analysis of the refined maxima cloud, organized as in Fig.~\ref{fig:minima_axis}.}
\label{fig:maxima_axis}
\end{figure}

\begin{figure}[t]
\centering
 \includegraphics[width=\textwidth]{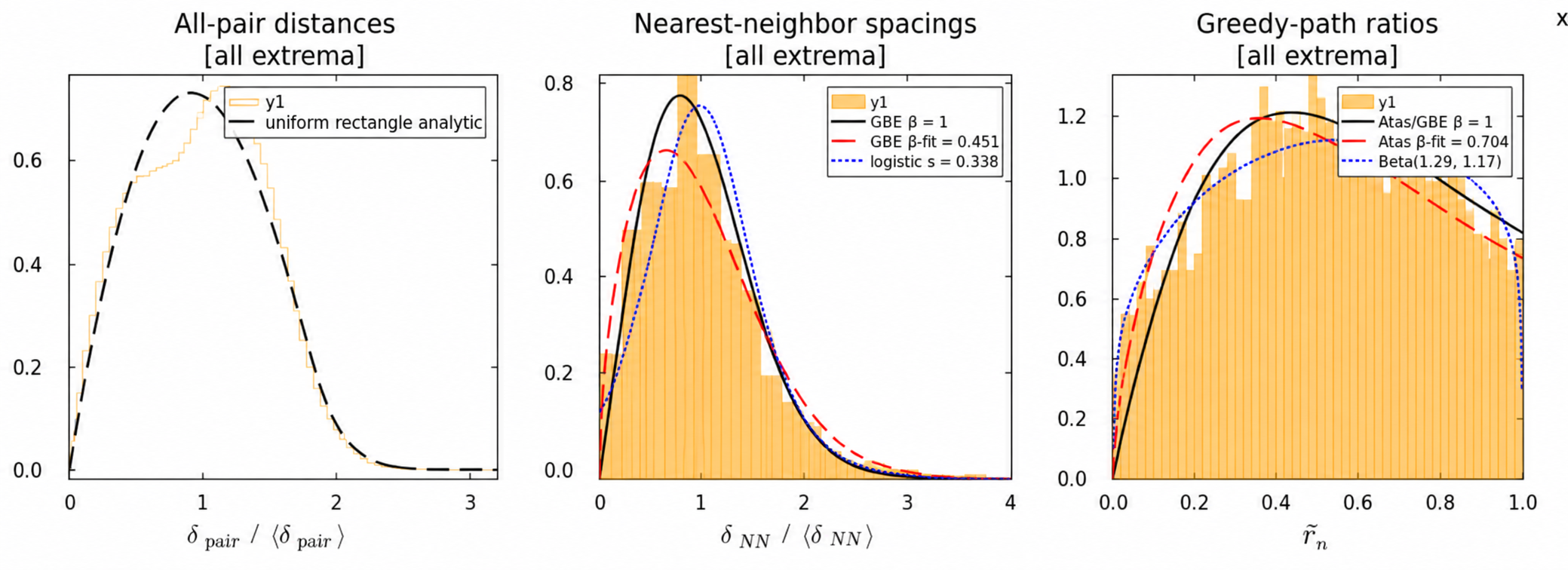}
\caption{Axis-rescaled, non-density-unfolded analysis of the refined combined cloud of all extrema, organized as in Fig.~\ref{fig:minima_axis}.}
\label{fig:all_axis}
\end{figure}

\subsection{Global geometry from all-pair distances}

The first observation, and it is a nontrivial one, is that the all-pair histograms in Figs.~\ref{fig:minima_axis}--\ref{fig:all_axis} lie rather close to the exact uniform-rectangle curve  once the trivial anisotropy between the $x$ and $t$ directions has been removed. This is true for minima, maxima, and the combined cloud, despite the fact that the total number of pairs is extremely large and therefore the histogram is sensitive to very small systematic distortions. For the three refined datasets shown, the total pair counts are $11{,}455{,}291$ for the minima cloud, $8{,}058{,}105$ for the maxima cloud, and $38{,}733{,}201$ for the combined cloud. These correspond to $N_-=4787$ minima, $N_+=4015$ maxima, and $N=8802$ extrema in the combined set. Since plotting all pair distances directly would be unnecessary and memory-intensive, the displayed histograms use a random sample of two million pairs in each case, which is already enough to suppress visible Monte Carlo noise.

What matters physically is not the sheer size of the sample but the shape of the deviation from the analytic rectangle law. The data do not simply reproduce the reference curve. In all three figures one sees a mild depletion at relatively small normalized pair distances, an enhancement in the vicinity of the modal region around the mean distance, and then a return toward the reference curve in the far tail. This is exactly the type of distortion one expects when the cloud has been made roughly isotropic by a global affine rescaling but has not been fully flattened in density. The crucial point is that the deviation is gentle rather than violent. The cloud is not concentrated on a thin set, nor does it exhibit a dramatic excess of very close pairs. Instead, it behaves globally like a fairly regular two-dimensional point set in a rectangle, with residual inhomogeneity superimposed on that global picture. This is important for the interpretation of the other two measures. It tells us that the nearest-neighbor and path statistics are not being extracted from a wildly singular geometry, but it also tells us that a true density-unfolding procedure remains conceptually necessary if one wishes to isolate a candidate universal local distribution.

A second important fact is that the all-pair histograms for maxima and minima are extremely similar. This is already suggestive. It implies that, at the scale probed by all pair distances, the sign of the extremum is not the primary organizing principle of the cloud. The maxima and minima are embedded in the same large-scale spacetime pattern and inherit nearly the same coarse geometry from it. The combined cloud therefore does not qualitatively alter the all-pair distribution. Instead, it mainly increases the sample size and slightly smooths the histogram. The lesson is that any difference between maxima and minima, if present, must be sought in more local or more ordered measures rather than in the global geometry of the cloud.

\subsection{Nearest-neighbor spacings and effective repulsion}

The nearest-neighbor spacing distributions tell a more refined story. For the minima cloud the Gaussian-$\beta$ fit gives
\begin{equation}\label{bNNm}
\beta_{\rm NN}^{(-)} \approx 0.4198,
\end{equation}
while the logistic fit gives
\begin{equation}
\sigma_{\rm log}^{(-)} \approx 0.3445.
\end{equation}
For the maxima cloud one finds
\begin{gather}
\beta_{\rm NN}^{(+)} \approx 0.6299, \\
\sigma_{\rm log}^{(+)} \approx 0.3268,
\end{gather}
and for the combined cloud
\begin{gather}\label{all}
\beta_{\rm NN}^{({\rm all})} \approx 0.4509, \\
\sigma_{\rm log}^{({\rm all})} \approx 0.3381.
\end{gather}
Several conclusions follow immediately. The first is that the GOE curve with $\beta=1$ is too strongly repulsive at small spacing and too sharply peaked near the mean spacing to describe the data. The effective repulsion present in the extremum cloud is therefore significantly weaker than the one-dimensional GOE benchmark. The second is that the fitted $\beta$ is nevertheless manifestly positive, so the cloud is not behaving like a one-dimensional Poisson spectrum either. The third is that maxima exhibit systematically stronger local repulsion than minima, although the difference is moderate rather than dramatic. The combined cloud lies between the two sign sectors and remains closer to the minima value than to the maxima value, which indicates that pooling the two sign sectors does not drive the system toward a GOE-like distribution.

The logistic fit is instructive here, not because it is theoretically privileged, but because it visibly captures a feature that the Gaussian-$\beta$ family struggles with in the present data. The histograms have a broad single hump centered near the mean spacing and they decay smoothly into the tail without the pronounced rigidity associated with a Wigner-like law. In that regime a logistic curve with width $\sigma \approx 0.34$ provides an economical description. This is precisely the kind of behavior that one expects in a genuine two-dimensional peak problem: local repulsion is present, but it is not naturally encoded by the same one-parameter family that describes ordered one-dimensional spectra. That the maxima, minima, and combined cloud all produce logistic widths within a few percent of one another is, in our view, one of the most robust pieces of information in the present dataset. It suggests that whatever local mechanism controls nearest-neighbor regularity is largely insensitive to the sign of the extrema and is instead set by the common underlying interference geometry of the correlator.

It is worth mentioning that in \cite{Bianchi:2025kna}  for the scattering of the quantum pinball, it was found out  that the fit of the  ``measured" spacings pdf to the logistic distribution was better than the one to the $G\beta E$.

At the same time one must resist the temptation to over-interpret the numerical value of $\beta_{\rm NN}$. As emphasized in the previous section, even a homogeneous two-dimensional Poisson process produces the GOE Wigner curve after mean normalization of nearest-neighbor spacings. The present data do not sit at $\beta=1$, but the general caution remains: nearest-neighbor repulsion in two dimensions is a geometric rather than uniquely spectral notion. The numerical value $\beta_{\rm NN}\approx 0.42$--$0.63$ should therefore be read as a compact measure of the softness of the local repulsion, not as a literal statement that the cloud belongs to a one-dimensional Gaussian ensemble with that Dyson index. From that point of view the main physics encoded in the middle panels of Figs.~\ref{fig:minima_axis}--\ref{fig:all_axis} is that the local operator quench generates an extremum cloud which is more regular than a featureless cluster but substantially less rigid than a GOE-like sequence.

\subsection{Greedy-path ratios and mesoscopic ordering}

The greedy-path ratio distributions are particularly revealing because they probe a level of organization that lies between nearest-neighbor geometry and fully global structure. For the minima cloud we find
\begin{gather}
\beta_r^{(-)} \approx 0.6271, \\
(a_-,b_-) \approx (1.298,1.216),
\end{gather}
for the maxima cloud
\begin{gather}
\beta_r^{(+)} \approx 0.7381, \\
(a_+,b_+) \approx (1.293,1.134),
\end{gather}
and for the combined cloud
\begin{gather}
\beta_r^{({\rm all})} \approx 0.7045, \\
(a_{\rm all},b_{\rm all}) \approx (1.293,1.175).
\end{gather}
The first striking feature is that these values are noticeably closer to $\beta=1$ than the nearest-neighbor exponents. This does not mean that the system becomes more random-matrix-like when one passes from measure II to measure III. Rather, it means that the path ratio statistic is probing a different layer of structure. The greedy path imposes a sequential organization on the point cloud, and once this is done the ratios of successive spacings become sensitive to the mesoscopic regularity of the cloud rather than only to the shortest local distances. The resulting distributions are broad, smooth, and concentrated away from the endpoints $\widetilde r=0$ and $\widetilde r=1$, which is precisely why the ordinary Beta family performs so well.

The ordinary Beta fits deserve special emphasis because they are not merely numerically convenient. When $a$ and $b$ are both slightly larger than unity, the Beta density has a broad interior maximum and mild suppression at the boundaries. This is exactly what one sees in the data. The three fitted parameter pairs lie in a narrow region of parameter space, with $a\simeq1.29$--$1.30$ and $b\simeq1.13$--$1.22$, and the resulting curves track the histogram more faithfully than the ABGR-type $\beta$-family across most of the interval. In physical terms, this tells us that the path statistic sees the extremum cloud as a relatively regular two-dimensional arrangement with a preferred mesoscopic spacing scale, rather than as a strictly one-dimensional sequence with universal adjacent-gap ratios. The fact that the fitted Beta parameters remain close across maxima, minima, and the combined set indicates that the sign of the extremum is a secondary datum at this level, even though the minima have a slightly smaller ABGR-type ratio exponent and a slightly larger fitted $b$ parameter.

It is also worth stressing that the path ratio statistic is inherently more model-dependent than the nearest-neighbor statistic, because it depends on the chosen ordering prescription and on the starting point of the path. In the present implementation the path begins at the leftmost point of the cloud, which is a natural and reproducible choice but by no means the only one. For that reason we do not present the path-ratio fits as a definitive universal fingerprint. We present them instead as evidence that the extremum cloud possesses a nontrivial mesoscopic organization which survives when one orders the cloud by a concrete geometric rule. The stability of the fitted Beta parameters across maxima, minima, and the full cloud suggests that this organization is not a fragile accident of one particular subset of extrema.

$\,$

Taken together, the three figures establish a coherent hierarchy of statements. First, at the global level, the extremum clouds look approximately like two-dimensional point sets in a rectangle once  the anisotropy is removed, although a clear residual inhomogeneity remains. Second, at the local level, the nearest-neighbor spacings are repulsive but considerably softer than the GOE Wigner law, with logistic fits offering a comparably good and in practice often better description. Third, at the mesoscopic level introduced by the greedy path, the ratio distributions are broader and more regular than the nearest-neighbor spacings and are described very naturally by ordinary Beta laws with parameters only slightly larger than unity. Fourth, all of these statements remain qualitatively stable when one passes from maxima to minima and then to the combined cloud, which strongly suggests that the sign of the extremum is not the leading organizing principle of the statistics.

What the figures do not yet establish is an unfolded universal law in the strong sense familiar from one-dimensional spectral statistics. The present figures use only the axis-rescaled metric and are therefore non-density-unfolded. This choice is deliberate. It displays the intrinsic geometry of the data with a minimal pre-processing step, and it makes it possible to see directly that the all-pair distribution is already rather close to the analytic rectangle law. At the same time, it leaves enough residual inhomogeneity that one must not mistake good phenomenological fits for proof of universality. In a genuine two-dimensional problem, density flattening can change the nearest-neighbor graph and hence the path itself. The purpose of a full unfolding study is therefore not to beautify the fits but to determine which parts of the statistics are stable under changes of coordinates and which parts are artifacts of the density profile. The present figures should be read in precisely that spirit: they identify the empirical laws that the unflattened extremum cloud seems to obey and thereby set the stage for a more systematic unfolding analysis.

\subsection{The dependence of $\beta$ on Euclidean smearing scale}

\begin{figure}[t!]
\centering
\includegraphics[width=1\textwidth]{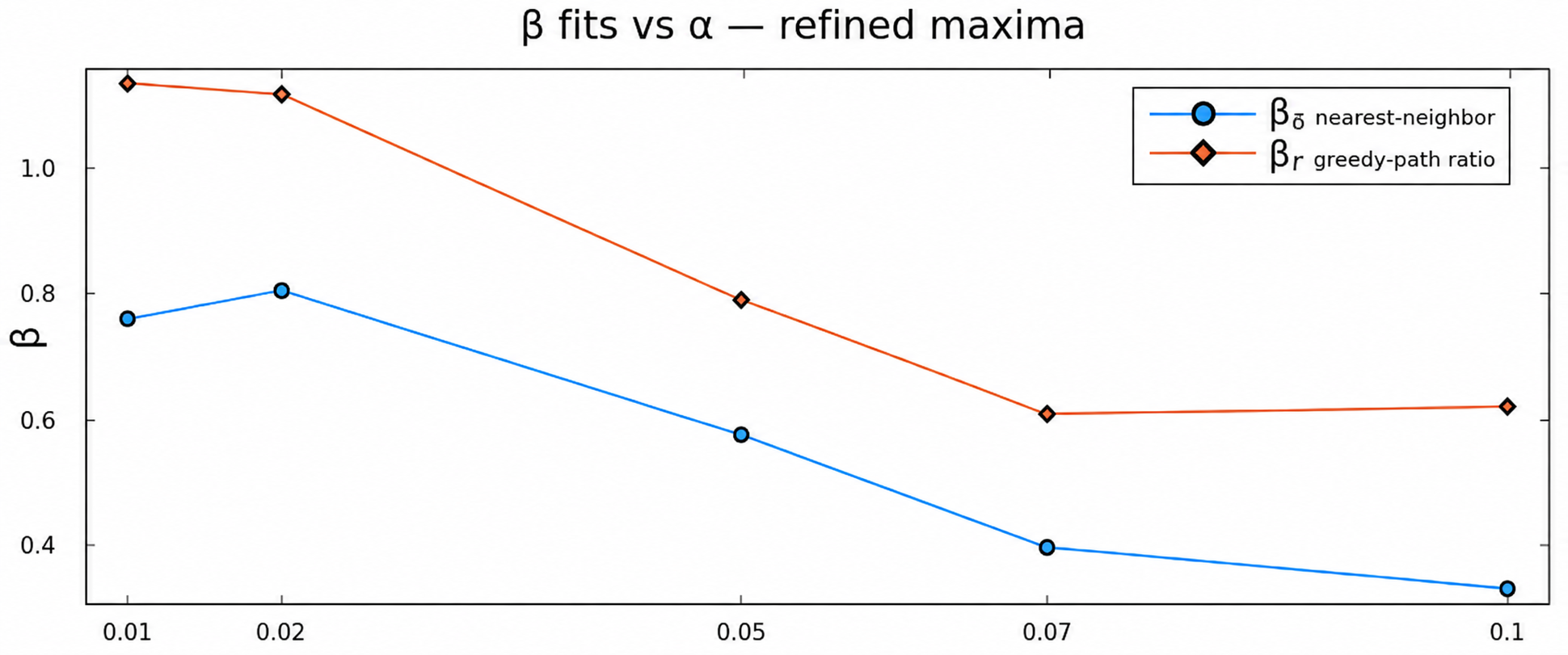}
\caption{Dependence of the fitted effective $\beta$ parameters on the Euclidean smearing scale $\alpha$ for the refined maxima cloud in the Neumann-Neumann interval sector, using the axis-rescaled metric and no density unfolding. The blue curve is the Gaussian-$\beta$ nearest-neighbor fit $\beta_\delta$, while the orange curve is the ABGR-type greedy-path ratio fit $\beta_r$.}
\label{figbetascanalpha}
\end{figure}
Finally, let us take present the results when the most of the picture comes closer to $\beta=1$ results and what intuition stands behind it. 
It is useful, before leaving the two-dimensional spacetime extremum cloud, to check how much of the preceding description is tied to the single Euclidean smearing scale used in the representative plots, since $\alpha$ is at once the regulator of the local operator insertion and the parameter that controls how much high-frequency mode content remains in the finite-volume interference pattern. Figure~\ref{figbetascanalpha} shows such a scan for the refined maxima in the Neumann-Neumann interval sector, with the same axis-rescaled metric used above and without any density unfolding, so that the change displayed in the figure should be read as a change in the physical smoothing of the correlator rather than as a change induced by a nonlinear flattening map.

The retained mode set is held fixed at 1656 Neumann-Neumann modes through the scan, while the number of maxima that enter the statistics decreases very rapidly from \(N=126188\) at $\alpha=0.01$ to \(N=41804\) at $\alpha=0.02$, then to \(N=7463\), \(3523\), and \(1610\) at $\alpha=0.05,0.07,0.1\). This sharp reduction is the expected finite-volume manifestation of Euclidean smearing, because increasing $\alpha$ damps the short-wavelength components that are responsible for a large fraction of the fine stationary structure in the $(x,t)$ plane.

The nearest-neighbor exponent $\beta_\delta$ displays a small hardening between the first two scan points, rising from approximately \(0.760\) to \(0.804\), but once the regulator is increased further it decreases rather steadily to \(0.575\), \(0.395\), and \(0.330\). The greedy-path ratio exponent $\beta_r$ follows the same qualitative softening, starting slightly above unity with \(1.134\) and \(1.117\) at the two smallest regulators and then falling to \(0.790\), \(0.608\), and \(0.621\), with the small upturn at the last point being well within the level of sensitivity expected when the surviving maxima cloud has become comparatively sparse. The important structural fact is that $\beta_r$ remains above $\beta_\delta$ throughout the scan, which is the same hierarchy already visible in the fixed-\(\alpha\) analysis and confirms that the greedy path continues to see a more ordered mesoscopic arrangement than the purely nearest-neighbor spacing distribution. This regulator dependence is also a useful check on the interpretation of the representative value $\alpha=0.05$. That point is rather an intermediate regime in which the cloud still contains enough maxima to give stable local and path statistics.

\section{The $2+1$-dimensional local quench model}
\label{3d}
Now let us consider 2+1-dimensional generalization of previous considerations, i.e. quench in the two-dimensional box, where the interval $[0,L]$ is replaced by the rectangular spatial domain $[0,L_x]\times [0,L_y]$ and the local-quench correlator becomes a function of the three variables $(x,y,t)$ rather than only of $(x,t)$. In this section we restrict to Neumann boundary conditions on all four sides of the box, because this is the case used in the numerical run described below and because it gives the cleanest higher-dimensional analogue of the interval NN sector. The theory is again the free massive scalar theory, now with action
\begin{equation}
S=\frac{1}{2}\int dt\int_0^{L_x}dx\int_0^{L_y}dy\,
\left[
(\partial_t\phi)^2-(\partial_x\phi)^2-(\partial_y\phi)^2-m^2\phi^2
\right],
\end{equation}
supplemented by the boundary conditions
\begin{gather}
\partial_x\phi(t,0,y)=0,\qquad \partial_x\phi(t,L_x,y)=0,\\
\partial_y\phi(t,x,0)=0,\qquad \partial_y\phi(t,x,L_y)=0.
\end{gather}
The corresponding one-dimensional Neumann modes in the two spatial directions are
\begin{gather}
u_0^{(x)}(x)=\frac{1}{\sqrt{L_x}},\qquad
u_n^{(x)}(x)=\sqrt{\frac{2}{L_x}}\cos\frac{n\pi x}{L_x},\qquad
k_n^{(x)}=\frac{n\pi}{L_x},\qquad n=1,2,\ldots,\\
u_0^{(y)}(y)=\frac{1}{\sqrt{L_y}},\qquad
u_\ell^{(y)}(y)=\sqrt{\frac{2}{L_y}}\cos\frac{\ell\pi y}{L_y},\qquad
k_\ell^{(y)}=\frac{\ell\pi}{L_y},\qquad \ell=1,2,\ldots.
\end{gather}
The two-dimensional Laplacian eigenfunctions are the tensor products
\begin{gather}
\Phi_{n\ell}(x,y)=u_n^{(x)}(x)u_\ell^{(y)}(y),\\
-\left(\partial_x^2+\partial_y^2\right)\Phi_{n\ell}(x,y)
=
q_{n\ell}^2\Phi_{n\ell}(x,y),\qquad
q_{n\ell}^2=\left(k_n^{(x)}\right)^2+\left(k_\ell^{(y)}\right)^2,
\end{gather}
with frequencies
\begin{equation}
\omega_{n\ell}=\sqrt{m^2+q_{n\ell}^2}.
\end{equation}
In terms of these modes the field operator is expanded as
\begin{equation}
\phi(t,x,y)=
\sum_{n,\ell\geq 0}
\frac{1}{\sqrt{2\omega_{n\ell}}}
\left(
a_{n\ell}e^{-i\omega_{n\ell}t}
+
a_{n\ell}^{\dagger}e^{i\omega_{n\ell}t}
\right)
\Phi_{n\ell}(x,y),
\end{equation}
where the zero mode is harmless in the present massive theory, but would have to be treated separately in the massless NN theory. The finite-temperature Euclidean two-point function in the box is therefore
\begin{equation}
G_{\beta_T}(\tau;{\bf r},{\bf r}')
=
\frac{1}{A}
\sum_{n,\ell\geq 0}
\frac{\Phi_{n\ell}({\bf r})\Phi_{n\ell}({\bf r}')}{2\omega_{n\ell}\left(1-e^{-\beta_T\omega_{n\ell}}\right)}
\left(
e^{-\omega_{n\ell}\tau}
+
e^{-\omega_{n\ell}(\beta_T-\tau)}
\right),
\end{equation}
where ${\bf r}=(x,y)$ and ${\bf r}'=(x',y')$. The local quench is produced by inserting the field at ${\bf r}_q=(x_q,y_q)$ with the same Euclidean smearing used in the interval problem,
\begin{equation}
|\Psi_\alpha(0)\rangle
=
\frac{e^{-\alpha H}\phi({\bf r}_q)|0\rangle}
{\langle 0|\phi({\bf r}_q)e^{-2\alpha H}\phi({\bf r}_q)|0\rangle^{1/2}},
\end{equation}
and the quantity whose extrema we study is the direct higher-dimensional analogue of the interval observable,
\begin{equation}
\mathcal{G}_{\rm loc}(t,x,y)=
\frac{2\,\mathrm{Re}\!\left[
G_{\beta_T}(\alpha+i t;{\bf r},{\bf r}_q)^\ast
G_{\beta_T}(\alpha+i t;{\bf r}_q,{\bf r}_q)
\right]}
{G_{\beta_T}(2\alpha;{\bf r}_q,{\bf r}_q)}.
\end{equation}
This formula is structurally identical to the one-dimensional local-quench expression, but the physics of the interference pattern is no longer the same, because the signal now propagates and reflects in two independent spatial directions and the extrema of $\mathcal{G}_{\rm loc}$ form a genuine three-dimensional spacetime cloud.

The analysis window is correspondingly promoted from a two-dimensional strip to a three-dimensional region in $(x,y,t)$, with a small core around the local insertion removed in order to suppress the short-distance region controlled mainly by the Euclidean regulator. We define
\begin{equation}
\Omega_3=
\left\{
(x,y,t)\,\big|\,
x_{\rm min}^{i}\leq x^{i}\leq x_{\rm max}^{i}, \qquad
(x-x_q)^2+(y-y_q)^2+t^2>r_{\rm core}^2
\right\}.
\end{equation}
The local maxima and minima are then the point sets
\begin{gather}
\mathcal{E}^{(3)}_+
=
\left\{
p_i^+=(x_i^+,y_i^+,t_i^+)\in\Omega_3
\,\big|\,
p_i^+ \ {\rm is\ a\ local\ maximum\ of}\ \mathcal{G}_{\rm loc}
\right\},\\
\mathcal{E}^{(3)}_-
=
\left\{
p_i^-=(x_i^-,y_i^-,t_i^-)\in\Omega_3
\,\big|\,
p_i^- \ {\rm is\ a\ local\ minimum\ of}\ \mathcal{G}_{\rm loc}
\right\},
\end{gather}
and the combined cloud is $\mathcal{E}^{(3)}=\mathcal{E}^{(3)}_+\cup\mathcal{E}^{(3)}_-$. Numerically, the raw extrema are first found by a $26$-neighbor test on the three-dimensional grid, and the accepted extrema are then refined by Newton iterations using analytic first and second derivatives of the mode sum. The Hessian test is also upgraded such that in the two-dimensional interval problem one can classify extrema using a determinant and trace condition, whereas in the present three-dimensional cloud the full $3\times3$ Hessian in the variables $(x,y,t)$ must have all eigenvalues negative for a maximum and all eigenvalues positive for a minimum.
\begin{figure}[t!]
    \centering
    \includegraphics[width=0.999\linewidth]{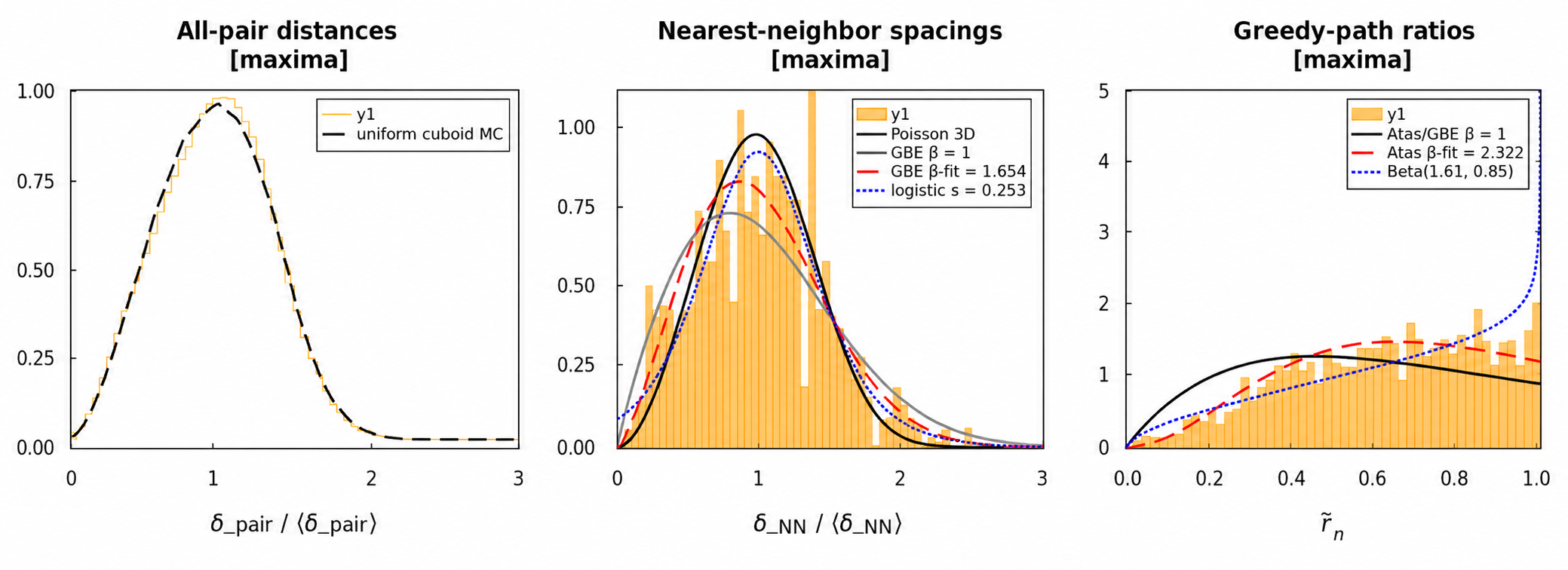}
\caption{Axis-rescaled analysis of the refined maxima cloud for the $2+1$-dimensional local quench in the two-dimensional square box with NN boundary conditions on all four sides. The left panel shows the histogram of all pair distances together with the uniform-cuboid Monte Carlo reference. The middle panel shows the nearest-neighbor spacing distribution together with the homogeneous Poisson reference in three dimensions, the Gaussian-$\beta$ comparison curves, and the logistic fit. The right panel shows the greedy-path ratio distribution together with the ABGR-type $\beta$ fit and the ordinary Beta fit.}
    \label{fig:maxima_3d_axis}
\end{figure}
The particular run shown in Fig.~\ref{fig:maxima_3d_axis} uses
\begin{gather}
m=10,\qquad
\beta_T=10000,\qquad
L_x=L_y=1,\qquad
A=1,\qquad
\alpha=0.05,\\
x_q=y_q=\frac{1}{2},\qquad
k_{\rm max}=350,\qquad
N_x=N_y=250,\qquad
N_t=500,\qquad
t\in[0.2,50],
\end{gather}
with an excluded core radius $r_{\rm core}=0.08$. The raw extremum extraction gives 
$N_+^{\rm raw}=4972,\qquad
N_-^{\rm raw}=5103,\qquad
N_{\rm all}^{\rm raw}=10075,$
while the Newton refinement, Hessian-eigenvalue classification, and deduplication step give
$
N_+^{\rm ref}=3403,\qquad
N_-^{\rm ref}=3463,\qquad
N_{\rm all}^{\rm ref}=6866.
$
The near equality between the number of surviving maxima and surviving minima is a useful diagnostic, because it indicates that the refined three-dimensional cloud is not dominated by an artificial preference for one sign sector and that the oscillatory pattern generated by the box quench produces a balanced population of positive and negative stationary structures.

One can see  that the all-pair distance histogram lies very close to the uniform-cuboid Monte Carlo reference. This agreement should be interpreted as a global statement about the geometry of the cloud rather than as a proof of local Poisson statistics. The three-dimensional extrema are not concentrated on a two-dimensional surface, a one-dimensional caustic, or a small number of coherent ridges; after removing only the trivial anisotropy between the coordinate directions, the maxima fill a metric volume whose coarse pair-distance distribution is already close to that of a uniform point set in a cuboid. This is the direct higher-dimensional analogue of the rectangle comparison in the interval problem, with the important difference that the reference geometry is now genuinely three-dimensional.

The nearest-neighbor data are more sensitive to local structure. For the refined maxima shown in Fig.~\ref{fig:maxima_3d_axis} the fitted values are
\begin{gather}
\beta_{\rm NN}^{(+)}\simeq 1.6541,\qquad
\sigma_{\rm log}^{(+)}\simeq 0.2530.
\end{gather}
 In three dimensions even an uncorrelated homogeneous point process has a geometrically induced suppression of very small nearest-neighbor distances, so the fitted Gaussian-$\beta$ parameter should not be read literally as a Dyson index. What the value $\beta_{\rm NN}^{(+)}\simeq1.65$ tells us is that the refined maxima cloud is locally more rigid than a soft two-dimensional interval cloud and has a well-developed microscopic spacing scale, while the logistic curve with width $\sigma_{\rm log}^{(+)}\simeq0.253$ captures the broad one-hump structure of the empirical distribution in a way that remains phenomenologically useful for multidimensional extrema.

The greedy-path ratio statistic gives an even sharper view of the mesoscopic organization of the same cloud. For the refined maxima one finds
\begin{gather}
\beta_r^{(+)}\simeq 2.3224,\qquad
(a_+,b_+)\simeq(1.613,0.850).
\end{gather}
The ABGR-type fit therefore moves to a rather large effective $\beta$, while the ordinary Beta fit develops an enhancement toward $\widetilde r=1$ through the parameter $b_+<1$. This is natural for a greedy walk through a nearly homogeneous three-dimensional point set, because consecutive nearest-unvisited-neighbor steps often have comparable lengths once the path has entered the bulk of the cloud. The right panel should therefore be read as evidence not merely for short-distance repulsion, but for a mesoscopic regularity of the extrema as a three-dimensional arrangement.

The remaining sign sectors from the same run give the following axis-rescaled, not density-unfolded, fitted values:
\begin{equation}
\begin{array}{c|c|c|c|c|c}
{\rm cloud} & N & \beta_{\rm NN} & \sigma_{\rm log} & \beta_r & (a,b) \\ \hline
{\rm raw\ maxima} & 4972 & 1.7381 & 0.2522 & 2.2008 & (1.613,0.892)\\
{\rm refined\ maxima} & 3403 & 1.6541 & 0.2530 & 2.3224 & (1.613,0.850)\\
{\rm raw\ minima} & 5103 & 1.4753 & 0.2726 & 2.2739 & (1.051,0.459)\\
{\rm refined\ minima} & 3463 & 1.2424 & 0.2751 & 1.9245 & (1.618,0.933)\\
{\rm raw\ all} & 10075 & 1.1942 & 0.2903 & 2.1962 & (1.216,0.613)\\
{\rm refined\ all} & 6866 & 0.9928 & 0.2936 & 1.8242 & (1.659,1.015)
\end{array}
\end{equation}
The higher-dimensional quench displays stronger local rigidity in the maxima sector, a softer but still nontrivial structure in the minima and combined sectors, and a path-ratio distribution that is naturally described by an ordinary Beta law rather than by a literal one-dimensional random-matrix ratio law. This is precisely the pattern one would expect if the higher-dimensional local quench is not simply reproducing spectral RMT in disguise, but is instead generating a genuinely multidimensional chaotic extremum geometry whose statistics mix wave-interference physics, finite-volume reflections, and the intrinsic geometry of point processes in three dimensions.

\section{All-pair  extrimum points spatial form factor}

Now let us define our final measure motivated by the spatial form factor of point patterns\cite{Massaro:2024xfd}. The observables discussed above use the extremum cloud directly in configuration space: the all-pair distance distribution probes its global geometry, the nearest-neighbor distribution probes its local regularity, and the greedy path probes an ordering-dependent mesoscopic structure. The spatial form factor gives a complementary representation of the first of these objects. It does not start from an ordering of the extrema and it does not isolate nearest neighbors. Instead, it takes all pairwise distances and performs an even Fourier transform of their empirical distribution. In this sense it is the natural point-cloud analogue of a spectral form factor, but with the role of level spacings replaced by metric distances between extrema.

Let $\mathcal{E}=\{p_i\}_{i=1}^N$ be any of the extremum clouds considered above, where the points may lie on a one-dimensional section, in the two-dimensional $(x,t)$ plane, or in the three-dimensional $(x,y,t)$ spacetime box. We denote by
\begin{equation}
d_{ij}=d(p_i,p_j)
\end{equation}
the corresponding metric distance, after the same coordinate rescaling or unfolding prescription used for the spacing observables. The all-pair spatial form factor is then defined as
\begin{equation}
E_xFF_N(k)= \frac{1}{N^2}\Re\left\langle \sum_{i,j=1}^{N} e^{ik d_{ij}}  \right\rangle
=\frac{1}{N^2}
\left\langle \sum_{i,j=1}^{N}
\cos\!\left(k\,d_{ij}\right).\right\rangle
\end{equation}
where $\left\langle. \right\rangle$ stands for avareging over the different configurations of $\mathcal{E}$. Here we are assume that the summation is taken over the total points set from the prescribed class (max, min or all extrema).
Equivalently, separating the diagonal terms $i=j$ from the off-diagonal pairs gives
\begin{equation}
E_xFF_N(k)=
\frac{1}{N}
+
\frac{2}{N^2}
\sum_{1\leq i<j\leq N}
\cos\!\left(k\,d_{ij}\right).
\end{equation}
This form is useful because it makes the large-$k$ plateau transparent. If the off-diagonal cosine phases decorrelate, the second term averages to zero and the form factor approaches the diagonal value $1/N$. It is also useful to introduce the normalized all-pair average
\begin{equation}
I_N(k)=
\frac{2}{N(N-1)}
\sum_{1\leq i<j\leq N}
\cos\!\left(k\,d_{ij}\right),
\end{equation}
so that
\begin{equation}
E_xFF_N(k)=
\frac{1}{N}
+
\left(1-\frac{1}{N}\right)I_N(k).
\end{equation}
Thus $K_N(0)=1$ by construction, while the plateau is fixed by the number of extrema and not by any fitted parameter.

In the numerical plots below we use the mean-normalized pair distance
\begin{equation}
\widehat d_{ij}=
\frac{d_{ij}}{\langle d\rangle_{\rm pair}},
\qquad
\langle d\rangle_{\rm pair}
=
\frac{2}{N(N-1)}
\sum_{1\leq i<j\leq N}d_{ij},
\end{equation}
and therefore the plotted momentum variable is the dimensionless conjugate variable to $\widehat d$. To avoid overloading the notation, we keep the symbol $k$ in the figure, but in this normalization $k$ should be read as $k\langle d\rangle_{\rm pair}$. In practice, the estimator used in the code is
\begin{equation}
E_xFF_N(k)=
\frac{1}{N}
+
\left(1-\frac{1}{N}\right)
\frac{1}{M}
\sum_{\alpha=1}^{M}
\cos\!\left(k\,\widehat d_{\alpha}\right),
\end{equation}
where $\{\widehat d_\alpha\}_{\alpha=1}^{M}$ is either the full list of normalized off-diagonal pair distances or a uniform random sample of that list when the number of pairs is too large. The reference curve is computed in exactly the same way, but with the extrema replaced by independent uniform points in the corresponding reference box: an interval for the one-dimensional section, a rectangle for the two-dimensional quench cloud, and a cuboid for the three-dimensional quench cloud. The diagonal contribution is always restored using the actual number $N$ of extrema in the empirical cloud, so that the reference and the data have the same plateau $1/N$.

\begin{figure}[t!]
    \centering
    \includegraphics[width=0.999\linewidth]{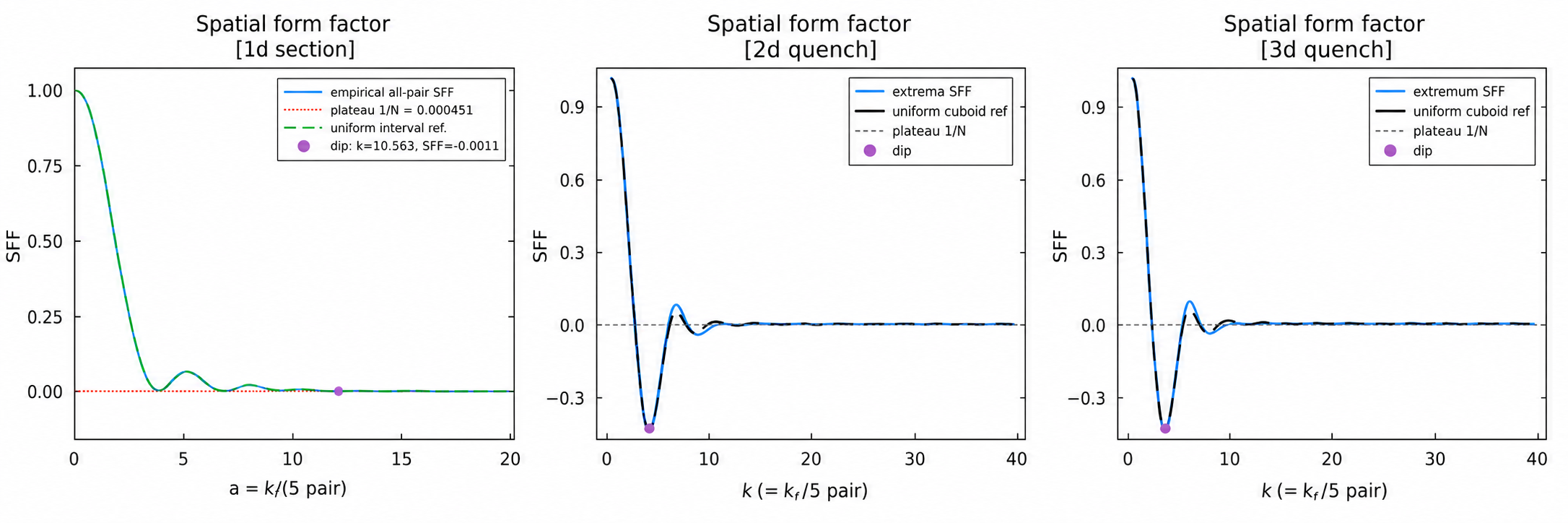}
\caption{All-pair spatial form factor for three versions of the local-quench extremum problem. The left panel shows a one-dimensional section of the two-dimensional quench data together with the uniform-interval reference. The middle panel shows the two-dimensional quench cloud together with the corresponding uniform-box reference. The right panel shows the three-dimensional quench cloud together with the uniform-cuboid reference. In all panels the form factor is computed from mean-normalized all-pair distances, starts at $K_N(0)=1$, and approaches the diagonal plateau $1/N$ after the off-diagonal cosine transform has dephased.}
    \label{fig:sff_triptych}
\end{figure}

The first panel of Fig.~\ref{fig:sff_triptych} is the most elementary benchmark. It takes a one-dimensional section of the two-dimensional local-quench data and compares its all-pair form factor to the uniform-interval reference. The agreement is almost exact over the entire displayed range. The form factor begins at unity and then shows a rapid decline as the all-pair phases dephase. In this one-dimensional case it is better not to call this feature a dip: the curve mainly describes the transition from the initial value $K_N(0)=1$ to the diagonal plateau of order $1/N$. The characteristic scale of this decline ends at  around
\begin{equation}
k_{\rm dec}\simeq 10.563,
\end{equation}
%which is close to the uniform-interval dephasing value $10\pi/3$.
This means that the one-dimensional section behaves almost like a uniform set of points on an interval. Thus, in the first panel the spatial form factor is mostly measuring the global interval geometry of the extrema, rather than a new random-matrix or chaotic correlation. It is therefore much less sensitive to the local repulsion and path-ordering effects seen in the nearest-neighbor and greedy-path statistics.
%%\sout{{\bf what does this mean}}

The middle panel shows the same construction for the two-dimensional quench cloud. Here the all-pair distances are the metric distances between extrema in the axis-rescaled $(x,t)$ plane, and the reference is the corresponding uniform box with the same effective global geometry. The empirical curve again follows the reference closely. The initial decay, the negative correlation hole, the first positive overshoot, and the subsequent damping are all reproduced by the uniform-box comparison\cite{Massaro:2024xfd}. This should be interpreted carefully. It does not mean that the two-dimensional extremum cloud is locally Poissonian, and it does not replace the nearest-neighbor or greedy-path analysis. Rather, it says that the dominant contribution to the all-pair form factor is the same dominant contribution already visible in the all-pair distance histogram: after the elementary axis rescaling, the cloud has a global pair-distance law close to that of a uniform point set in a bounded two-dimensional region.

The right panel repeats the analysis for the three-dimensional quench cloud in $(x,y,t)$. The reference is now a uniform cuboid, and the agreement remains striking. The first dip is deeper than in the one-dimensional section and the subsequent oscillations damp rapidly toward the plateau, but the empirical curve and the cuboid reference remain essentially locked together over the plotted range. This is precisely what one expects from the previous all-pair histogram in Fig.~\ref{fig:maxima_3d_axis}. The three-dimensional maxima cloud fills an effective metric volume with very little visible distortion at the level of global pair distances, and the spatial form factor simply expresses the same fact in the conjugate variable. The higher-dimensional curve is therefore not evidence for a new random-matrix universality by itself; it is evidence that the global geometry of the extremum cloud is already very close to the appropriate uniform-box geometry.

The main lesson of Fig.~\ref{fig:sff_triptych} is therefore somewhat different from the lesson of the nearest-neighbor and greedy-path figures. The form factor is an extremely efficient diagnostic of global filling. It confirms that the one-dimensional section is close to a uniform interval, that the two-dimensional quench cloud is close to a uniform rectangle after axis rescaling, and that the three-dimensional quench cloud is close to a uniform cuboid\cite{Massaro:2024xfd}. At the same time, precisely because it is built from all pairs, it is dominated by the same coarse geometry that controls the all-pair distance distribution. It is therefore less sensitive to the local stiffness of the maxima sector or to the endpoint enhancement of the greedy-path ratios. In the hierarchy of observables used in this paper, the spatial form factor should be viewed as the Fourier-space companion of the all-pair histogram: it verifies the large-scale metric support of the extremum cloud, while the nearest-neighbor and path-ratio statistics remain necessary for diagnosing the local and mesoscopic organization that is not visible in a purely global distance transform.
This interpretation also explains why the three panels look more similar than the corresponding nearest-neighbor statistics. Once the pair distances are normalized by their mean, the leading scale dependence is removed and the remaining form factor is controlled mainly by the shape of the effective support. The interval, rectangle, and cuboid references therefore organize the data almost completely. What survives as a genuinely dynamical question is not whether the all-pair form factor has a dip and a plateau, since even a uniform point process in a bounded region has such a structure, but whether the deviations from the uniform-box reference persist under density unfolding and whether they correlate with the softer or stiffer nearest-neighbor and path-ratio behavior found above. In this more conservative sense the spatial form factor closes the numerical story consistently: the local quench generates extremum clouds that are globally well distributed in their natural metric boxes, while their nontrivial chaotic content is carried by the residual local and mesoscopic correlations sitting on top of that global geometry.

\section{ Summary and open questions}
 The goal of the present paper was to quantify the erratic behavior of quenched correlation functions when they are treated as genuine functions of both space and time, rather than as one-dimensional time traces or spatial profiles.  For this purpose we implemented, in the setting of a local quench of a free massive scalar field theory on a finite interval, the multidimensional extremum-statistics framework introduced in \cite{Bianchi:2025kna}.  The first step was to compute the vacuum-subtracted two-point correlation function, displayed for a representative set of parameters in Fig.~\ref{fig:figG}.  Already at the level of this plot one observes a complicated spatiotemporal interference pattern, produced by the localized insertion, the finite-volume mode spectrum, and the repeated reflections from the boundaries, and it is this pattern which provides the raw material for the statistical analysis.

We then extracted the point clouds formed by the local maxima, by the local minima, and by the union of the two sets.  These clouds were analyzed through the same three complementary diagnostics used in \cite{Bianchi:2025kna}.  The first diagnostic was the all-pair distance distribution.  For each of the three extremum sets the corresponding histogram was found to be close, up to small but visible systematic deviations, to the distance distribution of a uniform point cloud in a rectangle after the elementary axis rescaling of the spacetime coordinates.  This shows that the extrema are not concentrated on a lower-dimensional structure or on a few preferred caustic curves, but instead fill the analysis window in a globally rather homogeneous way, with the remaining deviations reflecting the fact that no full density unfolding has yet been performed.

The second diagnostic was the nearest-neighbor spacing distribution.  After normalizing the spacings by their mean value, we compared the resulting probability density functions both with the Gaussian-$\beta$ ensemble spacing family and with the logistic distribution used as an effective comparison curve in multidimensional point-cloud problems.  The best-fit values of the effective repulsion parameter $\beta$ and of the logistic width $\sigma$ are given in \eqref{bNNm}--\eqref{all}.  In contrast with the behavior reported in \cite{Bianchi:2025kna} for the quantum pinball and related examples, the present local-quench data show that the Gaussian-$\beta$ fits are competitive with, and in the cases displayed here better than, the logistic fits.  This should not be interpreted as a literal identification of the two-dimensional extremum cloud with a one-dimensional random-matrix spectrum, but rather as evidence that the local extrema of the quenched correlator possess a definite short-distance repulsion whose effective strength is captured efficiently by the Gaussian-$\beta$ family.

The third diagnostic was provided by the greedy-path spacings and their adjacent ratios.  Since a multidimensional cloud has no canonical ordering, the greedy path supplies a concrete geometric prescription for converting the cloud into an ordered sequence, and the resulting ratio statistic probes a mesoscopic layer of organization which is not visible in the all-pair or nearest-neighbor data alone.  We compared the folded ratio histograms both with the ABGVV formula and with the ordinary Beta distribution on the unit interval.  The corresponding optimal values of the effective ratio parameter $\beta_r$ and of the Beta-distribution parameters $(a,b)$ show that the greedy-path statistic is again controlled by a broad, smooth distribution rather than by a rigid one-dimensional spectral law.  Taken together, the all-pair, nearest-neighbor, and greedy-path analyses indicate that the local quench produces an extremum geometry which is globally close to a uniform spacetime point cloud, locally repulsive, and mesoscopically ordered, while still retaining the geometric character of a genuinely multidimensional problem.
   . 

\subsection{What remains to be done}

Several steps remain before one can claim a mature and fully unfolded theory of extremum statistics for local quenches on an interval. The first is a systematic comparison of several two-dimensional unfolding prescriptions, including $x$-first and $t$-first Rosenblatt maps, marginal-CDF unfolding, and local-density metric rescalings. The second is a boundary-condition scan across the exact DD, NN, DN, and ND sectors, since the boundary reflections reorganize the spacetime pattern in a way that may affect the point cloud even when the gross statistics look similar. The third is a convergence study that reports not only fitted parameters but also tile-by-tile density diagnostics in the unfolded variables, so that the quality of the unfolding can be assessed directly rather than indirectly from the quality of a subsequent fit. The fourth is a study of path dependence, in which the initial point of the greedy path is varied systematically. Finally, the interpretation of the nearest-neighbor exponent should be placed on a firmer footing by comparing the data not only to repulsive $\beta$-families but also to local-density baselines that make the geometric origin of two-dimensional repulsion completely explicit.

These open tasks do not weaken the significance of the present results. On the contrary, they explain why the axis-rescaled figures shown in the main text are already useful. They identify the empirical distributions that any successful unfolded analysis must either reproduce or explain away. In that sense the current study should be viewed as the first precise map of the two-dimensional extremum geometry generated by a local quench on a finite interval, rather than as the final word on its universal statistics.

In addition there are plenty of other questions to be investigated. Here we enlist some of them:
\begin{itemize}
\item 
In this paper we have analyzed the properties of a free scalar theory with a local quench. One can introduce, as was done in \cite{Ageev:2025yiq} ,also global quenches. It will be interesting to inquire the chaotic behavior for such a quench. Moreover, one can analyze other free field theory with quenches like the Dirac fermion theory and Maxwell theory. Needless to say that quenches in interacting theories are also very interesting.
    \item 
In \cite{Bianchi:2024fsi} the one-dimensional chaotic SFF was shown to admit a ramp. On the other hand  the $d=1$ case in section section 6 does not have it. An understanding of the difference between the two behaviors is needed. 
\item 
The focus in this note was the extrema points. Generically  there are also saddle points and correspondingly Neumann lines and nodes\cite{band2016topological}. A natural question is  how do the latter fit the map to RMT. 
\item 
In \cite{Bianchi:2026jui} the multi-dimensional chaotic behavior was analyzed via  non-intersection curves in the scattering amplitude. It will be interesting to identify such curves also for the quenched systems. 

\end{itemize}
\section*{Acknowledgments}
We thank  M. Bianchi, M. Firrotta,  N. Shrayer and D. Weissman  for useful comments and discussions.
The work of JS was supported in part by a grant 01034816 titled ``String theory reloaded- from fundamental questions to applications'' of the ``Planning and budgeting committee''. 

\appendix

\section{Numerical implementation and  issues}

\subsection{Exact interval mode expansion}

The numerical evaluation of $\mathcal{G}_{\rm loc}(t,x)$ begins from the exact interval mode basis. In the exact Dirichlet and Neumann sectors the spatial eigenfunctions are known analytically, so the only numerical truncation enters through the maximum retained momentum $k_{\rm max}$ or, equivalently, the number of modes. The contribution of a mode of frequency $\omega_a$ to the locally excited state is weighted by the factor $u_a(x_q)e^{-\alpha \omega_a}$, and therefore modes for which $u_a(x_q)=0$ are present formally in the basis but absent dynamically from the quench profile. This simple observation matters in practice. For example, if $x_q=L/2$ in the DD sector, all even sine modes vanish at the insertion point and therefore decouple from the quench. The printed number of retained modes is therefore not always identical to the number of modes that are dynamically active in a given run. The physically relevant scale is set by the effective support of $e^{-\alpha\omega_a}$, not by the formal upper cutoff alone.

\subsection{Grid evaluation, differentiation, and extremum extraction}

Once the mode cache is built, the code evaluates $\mathcal{G}_{\rm loc}(t,x)$ on a dense rectangular grid in the $(x,t)$ plane. In the runs that underlie the figures shown in the main text, the grid contains $N_x=1000$ spatial points and $N_t=30000$ temporal points, which is sufficient to resolve the oscillatory pattern throughout the analysis window shown. Candidate extrema are first identified by an $8$-neighbor comparison on the grid. This step is intentionally redundant rather than minimal, because false positives are removed later. The raw candidates are then refined using analytic first and second derivatives of the correlator, obtained by differentiating the mode sum with respect to $x$ and $t$. A Newton step is used to move the candidate off the grid, and the resulting stationary point is accepted only if the Hessian has the correct signature, the determinant is sufficiently far from zero, and the refined point remains inside the local cell and inside the chosen analysis window. Near-duplicate points are then merged by a short-distance deduplication procedure. This is the key difference between a qualitative peak-finding routine and a quantitative extremum-statistics pipeline: the latter must treat the extremum positions themselves, not only the grid cells in which they were first detected.

\subsection{Metric coordinates and statistical estimators}

After the refined extrema are obtained, three statistical estimators are evaluated. The all-pair measure uses either the full set of pair distances or, when the pair count is too large, a controlled random sample of pairs. The nearest-neighbor spacings are computed either by brute force or by a nearest-neighbor data structure when available. The greedy-path statistic begins from a prescribed initial point and recursively attaches the nearest unvisited point. In the present draft the figures display the axis-rescaled metric rather than a full density unfolding. Concretely, the spatial and temporal coordinates are rescaled by the mean projected spacings along the two axes, which removes the trivial mismatch between the widths of the analysis window in the two directions while leaving the nonuniform density intact.

The fitted comparison curves are determined by straightforward likelihood maximization on one- or two-parameter grids. For nearest-neighbor spacings we fit both the Gaussian-$\beta$ family and a logistic distribution at fixed mean. For path ratios we fit both the folded ABGR-type $\beta$ family and an ordinary Beta distribution on $[0,1]$. For all-pair distances we compare directly to the exact distance law for two independent uniform points in a rectangle, after normalizing the distance by its mean. Since this last comparison is analytic, it provides a sharper diagnostic than a Monte Carlo baseline and makes it possible to distinguish a genuinely nonuniform cloud from one that merely suffers from finite histogram noise.

\subsection{Convergence and scales}

Although the examples shown in the main text use a rather conservative cutoff, it is useful to record the hierarchy of scales that controls numerical convergence. The ultraviolet regulator $\alpha$ sets the effective frequency scale through the factor $e^{-\alpha\omega_a}$, so that one generically needs $k_{\rm max}\gg \alpha^{-1}$ rather than an arbitrarily large basis. The temporal grid spacing $\Delta t$ must be fine enough to resolve the oscillatory modes that survive this damping. In other words, the mode cutoff and the time-grid resolution cannot be chosen independently if one wants a quantitatively controlled extremum analysis. In practice one should monitor the stability of the extremum counts, the all-pair histogram, the fitted nearest-neighbor parameters, and the fitted path-ratio parameters under simultaneous changes in $k_{\rm max}$, $N_x$, $N_t$, and the analysis window. A particularly important check, because it probes the reliability of the extremum locations themselves, is the stability of the results under changes of the Newton tolerance and of the deduplication threshold.

%\begin{thebibliography}{}

\bibliographystyle{JHEP}
%\end{thebibliography}
%\bibliography{local}

\begingroup
\renewcommand{\emph}[1]{#1}      % remove italics/emphasis
\renewcommand{\bfseries}{\mdseries} % remove bold volumes
\renewcommand{\ttfamily}{\normalfont} % remove typewriter arXiv numbers, optional
\bibliography{local}

@article{Bianchi:2025kna,
    author = "Bianchi, Massimo and Firrotta, Maurizio and Sonnenschein, Jacob and Weissman, Dorin",
    title = "{Multi-dimensional chaos I: Classical and quantum mechanics}",
    eprint = "2510.03007",
    archivePrefix = "arXiv",
    primaryClass = "hep-th",
    month = "10",
    year = "2025"
}

@article{Massaro:2024xfd,
    author = "Massaro, Matteo and del Campo, Adolfo",
    title = "{Spatial form factor for point patterns: Poisson point process, Coulomb gas, and vortex statistics}",
    eprint = "2410.04816",
    archivePrefix = "arXiv",
    primaryClass = "cond-mat.stat-mech",
    doi = "10.1103/PhysRevResearch.7.023107",
    journal = "Phys. Rev. Res.",
    volume = "7",
    number = "2",
    pages = "023107",
    year = "2025"
}

@article{Bianchi:2026jui,
    author = "Bianchi, Massimo and Firrotta, Maurizio and Sonnenschein, Jacob and Weissman, Dorin",
    title = "{Multi-dimensional chaos II: String scattering amplitudes, curve repulsion, and RMT}",
    eprint = "2606.24490",
    archivePrefix = "arXiv",
    primaryClass = "hep-th",
    month = "6",
    year = "2026"
}

@article{band2016topological,
  title={Topological Properties of Neumann Domains},
  author={Band, Ram and Fajman, David},
  journal={Annales Henri Poincar{\'e}},
  volume={17},
  number={9},
  pages={2379--2407},
  year={2016}
}

@article{ABGVV,
  author =       "Y. Y. Atas and E. Bogomolny and O. Giraud and E. Vivo and P. Vivo",
  title =        "{Joint probability densities of level spacing ratios in random matrices}",
  journal =      "J. Phys. A: Math. Theor.",
  volume =       "46",
  number =       "",
  pages =        "",
  year =         "2013",
  DOI =          "http://dx.doi.org/10.1002/andp.19053221004"
}

@article{Ageev:2025iiy,
    author = "Ageev, Dmitry S. and Bykov, Vladimir A.",
    title = "{From confinement to chaos in AdS/CFT via nonequilibrium local states}",
    eprint = "2507.22999",
    archivePrefix = "arXiv",
    primaryClass = "hep-th",
    doi = "10.1103/x8r7-sbrt",
    journal = "Phys. Rev. D",
    volume = "113",
    number = "4",
    pages = "046014",
    year = "2026"
}

@article{Bianchi:2022mhs,
    author = "Bianchi, Massimo and Firrotta, Maurizio and Sonnenschein, Jacob and Weissman, Dorin",
    title = "{Measure for Chaotic Scattering Amplitudes}",
    eprint = "2207.13112",
    archivePrefix = "arXiv",
    primaryClass = "hep-th",
    doi = "10.1103/PhysRevLett.129.261601",
    journal = "Phys. Rev. Lett.",
    volume = "129",
    number = "26",
    pages = "261601",
    year = "2022"
}

@article{Bianchi:2024fsi,
    author = "Bianchi, Massimo and Firrotta, Maurizio and Sonnenschein, Jacob and Weissman, Dorin",
    title = "{From spectral to scattering form factor}",
    eprint = "2403.00713",
    archivePrefix = "arXiv",
    primaryClass = "hep-th",
    reportNumber = "ITCP-IPP-2024/3",
    doi = "10.1007/JHEP06(2024)189",
    journal = "JHEP",
    volume = "06",
    pages = "189",
    year = "2024"
}

@article{Bianchi:2023uby,
    author = "Bianchi, Massimo and Firrotta, Maurizio and Sonnenschein, Jacob and Weissman, Dorin",
    title = "{Measuring chaos in string scattering processes}",
    eprint = "2303.17233",
    archivePrefix = "arXiv",
    primaryClass = "hep-th",
    doi = "10.1103/PhysRevD.108.066006",
    journal = "Phys. Rev. D",
    volume = "108",
    number = "6",
    pages = "066006",
    year = "2023"
}

@article{Ageev:2025yiq,
    author = "Ageev, Dmitry S. and Pushkarev, Vasilii V.",
    title = "{Random matrix theory signatures in free field theory}",
    eprint = "2507.18746",
    archivePrefix = "arXiv",
    primaryClass = "hep-th",
    month = "7",
    year = "2025"
}

@article{Bohigas:1983er,
    author = "Bohigas, O. and Giannoni, M. J. and Schmit, C.",
    title = "{Characterization of chaotic quantum spectra and universality of level fluctuation laws}",
    doi = "10.1103/PhysRevLett.52.1",
    journal = "Phys. Rev. Lett.",
    volume = "52",
    pages = "1--4",
    year = "1984"
}

@article{Berry:1977wk,
    author = "Berry, M. V. and Tabor, M.",
    title = "{Level clustering in the regular spectrum}",
    doi = "10.1098/rspa.1977.0140",
    journal = "Proc. Roy. Soc. Lond. A",
    volume = "356",
    pages = "375--394",
    year = "1977"
}

@book{Mehta:book,
    author = "Mehta, M. L.",
    title = "{Random Matrices}",
    edition = "Third",
    publisher = "Elsevier",
    year = "2004",
    doi = "10.1016/S0079-8169(04)80091-6"
}

@book{Haake:2010fgh,
    author = "Haake, Fritz",
    title = "{Quantum Signatures of Chaos}",
    doi = "10.1007/978-3-642-05428-0",
    isbn = "978-3-642-26330-9, 978-3-642-05428-0",
    publisher = "Springer",
    address = "Berlin",
    series = "Springer Series in Synergetics",
    year = "2010"
}

@book{Forrester:book,
    author = "Forrester, Peter J.",
    title = "{Log-Gases and Random Matrices}",
    publisher = "Princeton University Press",
    address = "Princeton",
    series = "London Mathematical Society Monographs",
    volume = "34",
    doi = "10.1515/9781400835416",
    year = "2010"
}

@article{Dumitriu:2002,
    author = "Dumitriu, Ioana and Edelman, Alan",
    title = "{Matrix models for beta ensembles}",
    eprint = "math-ph/0206043",
    archivePrefix = "arXiv",
    doi = "10.1063/1.1507823",
    journal = "J. Math. Phys.",
    volume = "43",
    pages = "5830--5847",
    year = "2002"
}

@article{Atas:2013dis,
    author = "Atas, Y. Y. and Bogomolny, E. and Giraud, O. and Roux, G.",
    title = "{Distribution of the ratio of consecutive level spacings in random matrix ensembles}",
    eprint = "1212.5611",
    archivePrefix = "arXiv",
    primaryClass = "math-ph",
    doi = "10.1103/PhysRevLett.110.084101",
    journal = "Phys. Rev. Lett.",
    volume = "110",
    pages = "084101",
    year = "2013"
}

@article{Rosenhaus:2020tmv,
    author = "Rosenhaus, Vladimir",
    title = "{Chaos in the Quantum Field Theory S-Matrix}",
    eprint = "2003.07381",
    archivePrefix = "arXiv",
    primaryClass = "hep-th",
    doi = "10.1103/PhysRevLett.127.021601",
    journal = "Phys. Rev. Lett.",
    volume = "127",
    number = "2",
    pages = "021601",
    year = "2021"
}

@article{Gross:2021gsj,
    author = "Gross, David J. and Rosenhaus, Vladimir",
    title = "{Chaotic scattering of highly excited strings}",
    eprint = "2103.15301",
    archivePrefix = "arXiv",
    primaryClass = "hep-th",
    doi = "10.1007/JHEP05(2021)048",
    journal = "JHEP",
    volume = "05",
    pages = "048",
    year = "2021"
}

@article{Rosenhaus:2021xhm,
    author = "Rosenhaus, Vladimir",
    title = "{Chaos in a Many-String Scattering Amplitude}",
    eprint = "2112.10269",
    archivePrefix = "arXiv",
    primaryClass = "hep-th",
    doi = "10.1103/PhysRevLett.129.031601",
    journal = "Phys. Rev. Lett.",
    volume = "129",
    number = "3",
    pages = "031601",
    year = "2022"
}

@article{Hashimoto:2022bll,
    author = "Hashimoto, Koji and Matsuo, Yoshinori and Yoda, Takuya",
    title = "{Transient chaos analysis of string scattering}",
    eprint = "2208.08380",
    archivePrefix = "arXiv",
    primaryClass = "hep-th",
    doi = "10.1007/JHEP11(2022)147",
    journal = "JHEP",
    volume = "11",
    pages = "147",
    year = "2022"
}

@article{Savic:2024ock,
    author = "Savi\'c, Nikola and \v{C}ubrovi\'c, Mihailo",
    title = "{Weak chaos and mixed dynamics in the string S-matrix}",
    eprint = "2401.02211",
    archivePrefix = "arXiv",
    primaryClass = "hep-th",
    doi = "10.1007/JHEP03(2024)101",
    journal = "JHEP",
    volume = "03",
    pages = "101",
    year = "2024"
}

@article{Calabrese:2007mt,
    author = "Calabrese, Pasquale and Cardy, John",
    title = "{Quantum quenches in extended systems}",
    eprint = "0704.1880",
    archivePrefix = "arXiv",
    primaryClass = "cond-mat.stat-mech",
    doi = "10.1088/1742-5468/2007/06/P06008",
    journal = "J. Stat. Mech.",
    volume = "0706",
    pages = "P06008",
    year = "2007"
}

@article{Calabrese:2007rg,
    author = "Calabrese, Pasquale and Cardy, John",
    title = "{Entanglement and correlation functions following a local quench: a conformal field theory approach}",
    eprint = "0708.3750",
    archivePrefix = "arXiv",
    primaryClass = "quant-ph",
    doi = "10.1088/1742-5468/2007/10/P10004",
    journal = "J. Stat. Mech.",
    volume = "0710",
    pages = "P10004",
    year = "2007"
}

@article{Nozaki:2013wia,
    author = "Nozaki, Masahiro and Numasawa, Tokiro and Takayanagi, Tadashi",
    title = "{Holographic Local Quenches and Entanglement Density}",
    eprint = "1302.5703",
    archivePrefix = "arXiv",
    primaryClass = "hep-th",
    doi = "10.1007/JHEP05(2013)080",
    journal = "JHEP",
    volume = "05",
    pages = "080",
    year = "2013"
}

@article{Nozaki:2014hna,
    author = "Nozaki, Masahiro and Numasawa, Tokiro and Takayanagi, Tadashi",
    title = "{Quantum Entanglement of Local Operators in Conformal Field Theories}",
    eprint = "1401.0539",
    archivePrefix = "arXiv",
    primaryClass = "hep-th",
    doi = "10.1103/PhysRevLett.112.111602",
    journal = "Phys. Rev. Lett.",
    volume = "112",
    pages = "111602",
    year = "2014"
}

@article{Cotler:2016fpe,
    author = "Cotler, Jordan S. and Gur-Ari, Guy and Hanada, Masanori and Polchinski, Joseph and Saad, Phil and Shenker, Stephen H. and Stanford, Douglas and Streicher, Alexandre and Tezuka, Masaki",
    title = "{Black Holes and Random Matrices}",
    eprint = "1611.04650",
    archivePrefix = "arXiv",
    primaryClass = "hep-th",
    doi = "10.1007/JHEP05(2017)118",
    journal = "JHEP",
    volume = "05",
    pages = "118",
    year = "2017",
    note = "[Erratum: JHEP 09, 002 (2018)]"
}

@article{Liu:2018hlr,
    author = "Liu, Junyu",
    title = "{Spectral form factors and late time quantum chaos}",
    eprint = "1806.05316",
    archivePrefix = "arXiv",
    primaryClass = "hep-th",
    doi = "10.1103/PhysRevD.98.086026",
    journal = "Phys. Rev. D",
    volume = "98",
    number = "8",
    pages = "086026",
    year = "2018"
}

@article{Ageev:2022kpm,
    author = "Ageev, Dmitry S. and Belokon, Aleksandr I. and Pushkarev, Vasilii V.",
    title = "{From locality to irregularity: introducing local quenches in massive scalar field theory}",
    eprint = "2205.12290",
    archivePrefix = "arXiv",
    primaryClass = "hep-th",
    doi = "10.1007/JHEP05(2023)188",
    journal = "JHEP",
    volume = "05",
    pages = "188",
    year = "2023",
    note = "[Erratum: JHEP 12, 184 (2023)]"
}

@article{Heyl:2018otc,
    author = "Heyl, Markus and Pollmann, Frank and D{\'o}ra, Bal{\'a}zs",
    title = "{Detecting equilibrium and dynamical quantum phase transitions in Ising chains via out-of-time-ordered correlators}",
    eprint = "1801.01684",
    archivePrefix = "arXiv",
    primaryClass = "cond-mat.str-el",
    doi = "10.1103/PhysRevLett.121.016801",
    journal = "Phys. Rev. Lett.",
    volume = "121",
    number = "1",
    pages = "016801",
    year = "2018"
}

@article{Alba:2019scrambling,
    author = "Alba, Vincenzo and Calabrese, Pasquale",
    title = "{Quantum information scrambling after a quantum quench}",
    eprint = "1903.09176",
    archivePrefix = "arXiv",
    primaryClass = "cond-mat.stat-mech",
    doi = "10.1103/PhysRevB.100.115150",
    journal = "Phys. Rev. B",
    volume = "100",
    number = "11",
    pages = "115150",
    year = "2019"
}

@article{Balasubramanian:2019quench,
    author = "Balasubramanian, Vijay and Craps, Ben and De Clerck, Marine and Nguyen, K{\'e}vin",
    title = "{Superluminal chaos after a quantum quench}",
    eprint = "1908.08955",
    archivePrefix = "arXiv",
    primaryClass = "hep-th",
    doi = "10.1007/JHEP12(2019)132",
    journal = "JHEP",
    volume = "12",
    pages = "132",
    year = "2019"
}

@article{Das:2021otoc,
    author = "Das, Suchetan and Ezhuthachan, Bobby and Kundu, Arnab and Porey, Somnath and Roy, Baishali",
    title = "{Critical Quenches, OTOCs and Early-Time Chaos}",
    eprint = "2108.12884",
    archivePrefix = "arXiv",
    primaryClass = "hep-th",
    doi = "10.1007/JHEP07(2022)046",
    journal = "JHEP",
    volume = "07",
    pages = "046",
    year = "2022"
}

@article{Caputa:2025krylov,
    author = "Caputa, Pawel and Di Giulio, Giuseppe",
    title = "{Local Quenches from a Krylov Perspective}",
    eprint = "2502.19485",
    archivePrefix = "arXiv",
    primaryClass = "hep-th",
    reportNumber = "YITP-25-26",
    doi = "10.1007/JHEP07(2025)164",
    journal = "JHEP",
    volume = "07",
    pages = "164",
    year = "2025"
}
\endgroup
\end{document}